\documentclass[a4paper, 12pt, BCOR=2cm, headsepline, bibliography=totoc, index=totoc]{scrbook}
\usepackage{type1cm}
\usepackage{svnkw}
\usepackage{tabularx}
\usepackage{times}
\usepackage[slantedGreek]{mathptmx}
\usepackage[latin1]{inputenc}
\usepackage[english, ngerman]{babel}
\usepackage{graphicx}
\usepackage{booktabs}
\usepackage{float}
\usepackage{color}
\usepackage{amssymb}
\usepackage{amsmath}
\usepackage{theorem}
\usepackage{makeidx}
\usepackage{listings}
\usepackage{placeins}
\usepackage{multirow}
\usepackage{fancybox}
\usepackage{pbox}
\usepackage{csquotes}
\usepackage[newcentury]{myquotchap}
\usepackage{url}
\usepackage{xpatch}

\usepackage{pdfpages}

\usepackage[style=numeric, defernumbers=true, sorting=ydnt, maxbibnames=99, url=false, doi=false, isbn=false,sortlocale=nb_NO]{biblatex}
\DeclareBibliographyCategory{patents}
\addtocategory{patents}{patep,patus}

\DeclareBibliographyCategory{books}
\addtocategory{books}{KHZ16,HKZ15,HKZ14,zie10}

\DeclareBibliographyCategory{bookchapters}
\addtocategory{bookchapters}{KZH16,zst10}

\DeclareBibliographyCategory{journals}
\addtocategory{journals}{ZBB16,WRZT:2013,ZT09,ZT07}

\DeclareBibliographyCategory{conferences}
\addtocategory{conferences}{EWBTZ16, PZ6, BEZWT:2016, BZMT15,BEZT15,GRBZFTH15,DBLP:conf/fpl/Becher14, RAW:2014, FCCM:2014, Schmidt:2014:ANF:2554688.2554730, SZT14,FCCM:2013,  ZBTZ12,GYRT2012, FCCM:2012, KZF12,DBLP:conf/fpt/AngermeierZGT11,DBLP:conf/codes/WildermannRZT11,DBLP:conf/fpl/AngermeierZGT11, DBLP:conf/fpl/WildermannTZ11, ZWOWT11, TZ11, ZT10,ZBT10b, ZBT10,
 MBZ10,SZT2008,ZT2008, AISEC2007, AIS2007, ZT2006, ZAT2006,AHWZ06}

\DeclareBibliographyCategory{miscs}
\addtocategory{miscs}{DBLP:conf/fpt/ZiermannSMZAT11,AIS2009,ZT05}

\DeclareBibliographyCategory{important}
\addtocategory{important}{ZBB16,FCCM:2012,DBLP:conf/fpl/Becher14,BZMT15,DBLP:conf/fpt/AngermeierZGT11,FCCM:2014,GRBZFTH15}

\newcommand*{\addcat}{%
  \ifcategory{important}%
    {\color{chaptergrey}}%
    {}%
}
\renewbibmacro*{begentry}{\addcat}

\xpatchbibmacro{cite}
  {\printfield{labelnumber}}{\ifcategory{important}{{\printfield{labelnumber}*}}{\printfield{labelnumber}}}
  {}{\typeout{failed to patch cite:comp:comp macro}}

\xpatchbibmacro{cite:comp:end}
  {\printfield{labelnumber}}{\printfield{labelnumber}\addcat}
  {}{\typeout{failed to patch cite:comp:comp macro}}

\xpatchbibmacro{cite:comp:inset}
  {\printfield{labelnumber}}{\printfield{labelnumber}\addcat}
  {}{\typeout{failed to patch cite:comp:comp macro}}

\makeatletter
\def\url@leostyle{%
  \@ifundefined{selectfont}{\def\UrlFont{\sf}}{\def\UrlFont{\small\ttfamily}}}
\makeatother
\floatstyle{ruled}
\floatname{Listing}{\sffamily\bfseries Listing}
\newfloat{Listing}{tbp}{lis}[chapter]

\lstdefinelanguage{sparc}{
     keywords={nop, save, inc, ret, restore, cmp, bg, call, b}}

\usepackage[algoruled,algochapter,lined,linesnumbered]{algorithm2e}
\dontprintsemicolon
\DeclareSymbolFont{symbols}%
{OMS}{pzc}{m}{n}
\DeclareSymbolFontAlphabet{\mathcal}%
{symbols}

{\theoremheaderfont{\normalfont\sffamily\bfseries}

{\theorembodyfont{\normalfont}
}}


\newcommand{\habiltitle}[3]{
  \begin{titlepage}
         \title{\Huge{#1}}
               
  \author{\\[10mm]
          \large  \sffamily Der Technischen Fakult{\"a}t der\\
          \large \sffamily Friedrich-Alexander-Universit{\"a}t Erlangen-N{\"u}rnberg\\[10mm]
          \large \sffamily als\\[10mm]
          \sffamily \textbf{Habilitationsschrift}\\[10mm]
          \large \sffamily vorgelegt von\\[10mm]
          \sffamily\Large #2\\[16mm]
  }
  \date{\large  \sffamily Hamburg 2017}
  \thispagestyle{empty}
  \end{titlepage}  
  \lowertitleback{ \parbox{12cm}{\normalsize \sffamily  Als Habilitation genehmigt von der Technischen Fakult{\"a}t der Friedrich-Alexander-Universit{\"a}t Erlangen-N{\"u}rnberg}\\[1cm]
\parbox{\textwidth}{\normalsize \sffamily Tag der Einreichung: \dotfill 08. M{\"a}rz 2017\\
Erteilung der Lehrbef{\"a}higung (\emph{venia legendi}): \dotfill 13. Dezember 2017\\ [10 mm]
Fachmentorat: 
\begin{itemize}
\item Prof. Dr.-Ing. J{\"u}rgen Teich
\item  Prof. Dr. Klaus Meyer-Wegener
\item Prof. Dr.-Ing. Dietmar Fey
\end{itemize} %
 \quad \\ [5 mm]
Gutachter:
\begin{itemize}
\item Prof. Dr. Marco Platzner, Universit{\"a}t Paderborn
\item Prof. Dr. Oliver Diessel, University of New South Wales, Australia
\end{itemize}
}}

  \maketitle
}

\makeindex

\usepackage[hidelinks]{hyperref}

\begin{document}
\frontmatter
\habiltitle{Improving Reliability, Security, and Efficiency of Reconfigurable Hardware Systems}{Daniel Michael Ziener}{ \svnkw{LastChangedRevision}}
\selectlanguage{english}
\chapter*{Abstract} 
 
In this treatise,  my research on methods to improve efficiency, reliability, and security of reconfigurable hardware systems, i.e., FPGAs, through partial dynamic reconfiguration is outlined. The efficiency of reconfigurable systems can be improved by loading optimized data paths on-the-fly on an FPGA fabric. 
This technique was applied to the acceleration of SQL queries  for large database applications as well as for image and signal processing applications. The focus was not only on performance improvements and resource efficiency, but also the energy efficiency has been significantly improved. In the area of reliability, countermeasures against radiation-induced faults and  aging effects for long mission times were investigated and applied to SRAM-FPGA-based satellite systems. Finally, to increase the security of cryptographic FPGA-based implementations against physical attacks, i.e., side-channel and fault injection analysis as well as reverse engineering, it is proposed to transform static circuit structures into dynamic ones by applying dynamic partial reconfiguration. 

\tableofcontents
\clearpage
\mainmatter

\chapter{Introduction} \label{sec:intro}

\emph{Reconfigurable hardware systems} are able to implement a desired circuit structure according a given configuration. 
The research field of architectures of such reconfigurable hardware systems as well as algorithms and applications involving reconfigurable hardware systems is called \emph{Reconfigurable Computing}. 

The mainly used reconfigurable devices are \emph{Field Programmable Gate Arrays} (FPGAs) which belong to the family of \textit{PLDs} (Programmable Logic Devices). 
FPGAs are digital chips that can be programmed for implementing arbitrary digital circuits. 
This means that FPGAs have first to be programmed with a so called \emph{configuration} (often called a configuration bitstream) to set the desired behavior of the used functional elements of the FPGA.
FPGAs have a significant market segment in the microelectronics and, particularly in the embedded system area.  
For example, FPGAs are commonly used in network equipment, avionics\footnote{In an article from EETimes it is stated that "Microsemi already has over 1000 FPGAs in every Airbus A380".
(http://www.electronics-eetimes.com/?cmp\_id=7\&news\_id=222914228)}, 
automotive, automation, various kinds of test equipment, medical devices, just to name some application domains.

\section{Flexibility vs. Efficiency}

\begin{figure}[t] \centering 
\includegraphics[width=0.8\textwidth]{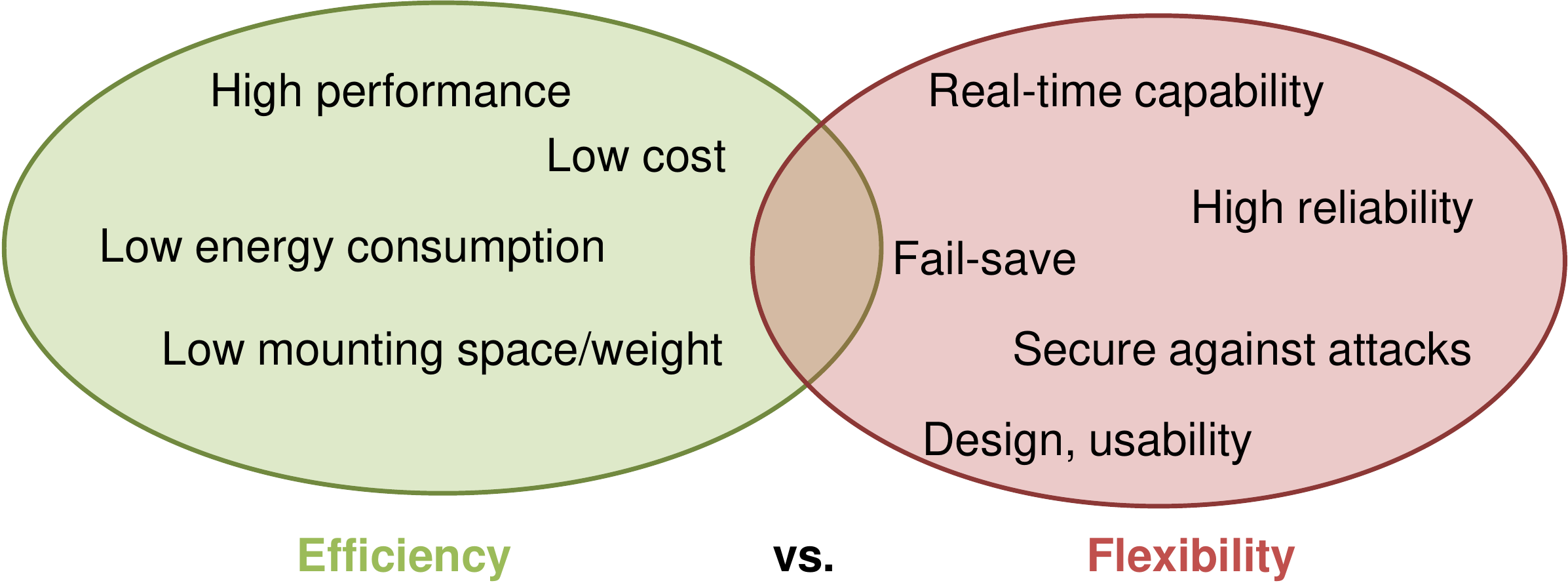} 
\caption{Several important properties of embedded systems which have to be considered during design and implementation are depicted. The properties are categorized into properties which need a flexible and properties which need an efficient implementation platform.} \label{fig:FlexVsEff-crop}
\end{figure}

\begin{figure}[t] \centering 
\includegraphics[width=0.9\textwidth]{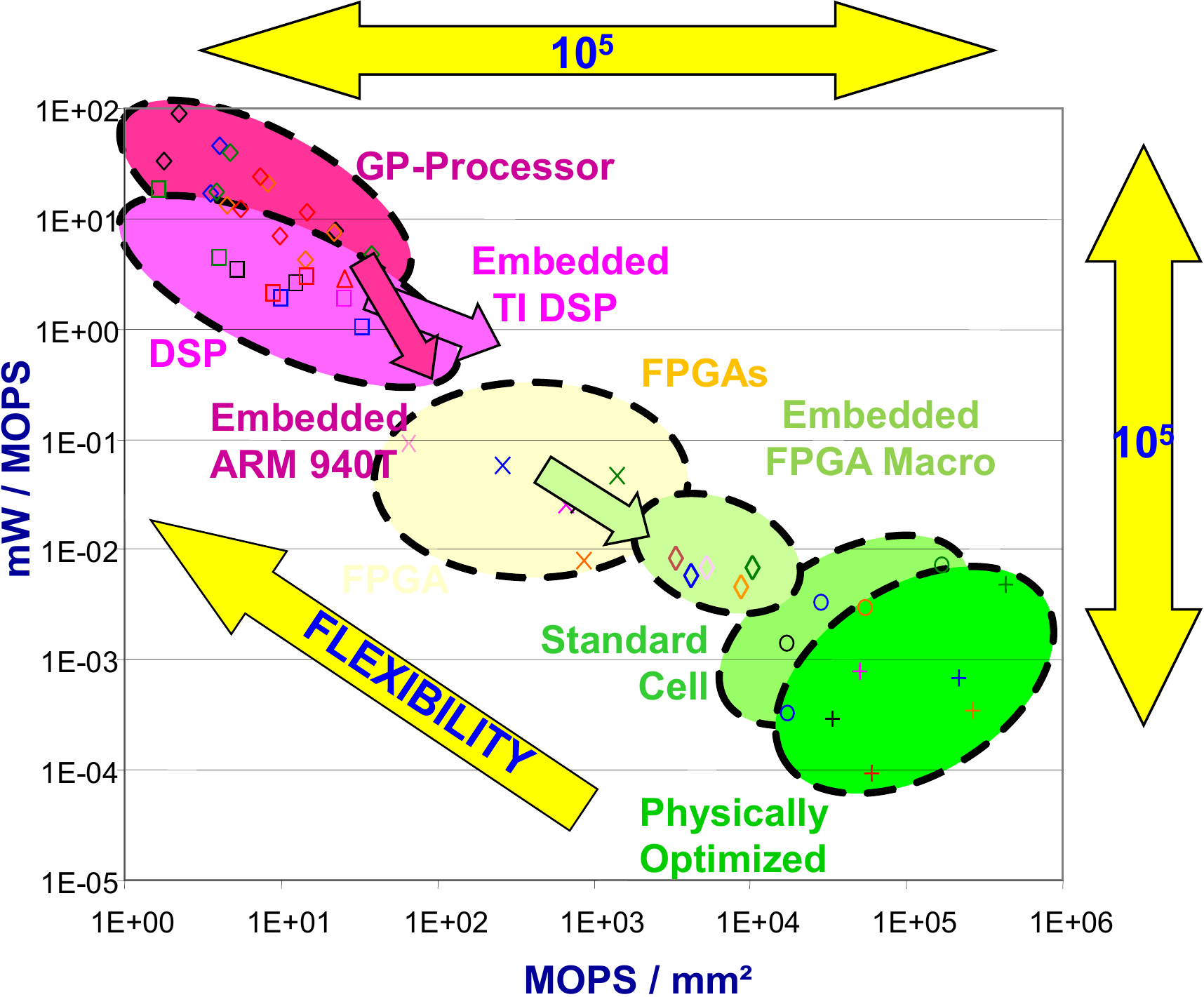} 
\caption{Energy and area efficiency of different implementations of sample applications. All entries are properly scaled to 130 nm CMOS technology. Furthermore, the increasing flexibility is depicted. Taken from \cite{Noll2010}.} \label{fig:noll-crop}
\end{figure}

One reason why FPGAs are so successful as an implementation platform for embedded systems is their combination of \emph{flexibility} and \emph{efficiency}. Efficiency is mostly defined as area efficiency in performance per area, e.g., MOPS (mega operations per second) per $mm^2$, or as energy efficiency in performance per power, e.g., MOPS per Watt. Figure \ref{fig:FlexVsEff-crop} shows some of the most important required or desirable properties of embedded systems. If we analyze the different properties, we will find out that some properties need an efficient implementation platform, e.g., \emph{low cost, low energy consumption, high performance, low mounting space} or \emph{weight}.
Whereas other properties need a flexible platform, either to adapt the implementation or the functionality. The adaption of the implementation is needed in order to react on different disturbances, e.g., to ensure the \emph{real-time capability} by using more resources,  to ensure \emph{reliability} and \emph{fail-save} properties by using redundant structures, or to be \emph{secure against attacks} by including attack-specific countermeasures during run time. On the other hand, the fast adaption of functionally is needed in order to react on user requests or external triggers which increases the \emph{usability} of the system. 
If we look at available implementation platforms (see Figure \ref{fig:noll-crop} from \cite{Noll2010}), we can identify very flexible, but inefficient implementation platforms, like general purpose processors. On the other hand, very efficient platforms, like physically optimized ASICs, using a fixed circuit structure and are, therefore, very inflexible. To find a suitable platform to implement an embedded system that is as well flexible as efficient, one could start from the most efficient platform and graudally increase the flexibility. The \emph{standard cell} approach has an increased flexibility in the design flow. However, the resulting circuits have still a fixed structure. A design style which allows to reconfigure a given circuit structure is provided by FPGAs. However, to leverage the full flexibility of FPGAs, a technique called \emph{dynamic partial reconfiguration} (DPR) has to be used (see Chapter \ref{sec:reconSys}). The combination of flexibility and efficiency makes FPGAs to a widely used and very successful platform for embedded systems. 

However, not only embedded systems are in focus of this treatise. Also the usage of FPGAs to implement flexible hardware accelerators for data centers is an important key aspect. 
While currently only a tiny share of FPGAs are used for data processing in data centers, this is likely going to change in the near future as FPGAs not only provide very high performance, but they are also extremely energy efficient computing devices.
This holds in particular when considering big data processing.
For example, Microsoft recently demonstrated a doubling of the ranking throughput of their Bing search engine by equipping 1,632 servers with FPGA accelerators which only added an extra 10\% in power consumption~\cite{putnam2014reconfigurable}.
In other words, Microsoft was able to improve the energy efficiency by 77\% while providing faster response times due to introducing FPGAs in their data centers.

\section{Reliability and Security}

Their success makes FPGAs also interesting for new safety- and security-critical applications and application fields. However, in these new operation sites, FPGAs have to deal with harsh environments, and the implemented systems are forced to guarantee a high reliability and/or security. Especially SRAM-based FPGAs have to deal with radiation-induced errors like single event effects, for example, in space missions, avionics, or in an attacker's laboratory who tries to get sensitive information out of the FPGA by using \emph{fault attacks} \cite{bi97}.

Moreover, FPGAs  as well as all other integrated circuits suffer from the negative side effects of  the advances in the underlying technology. Due to the ever shrinking transistor sizes, the digital foundation from which FPGAs are build upon is getting more and more unreliable. This leads to an increased sensitivity against radiation effects as well as an accelerated aging of the digital circuits.  The great challenge is to design reliable systems from unreliable components \cite{bor05}. In the past, the  need for additional functions to improve the reliability of a system through error monitoring and correction was only given for safety-critical systems. Examples are banking mainframes, control systems of nuclear plants, and chip cards. In the future, the need for reliability-preserving and -increasing techniques will also become substantial for  consumer products.

On the other hand, also security becomes more and more important for embedded systems. With the ongoing integration of  embedded systems into networks, security attacks on these systems arose. Also, the increased complexity of these systems increases the probability of errors which can be used to break into a system. A common objective for  attackers is sensitive data, which is stored inside a digital system. Physical attacks, where digital systems are physically penetrated to gather sensitive information, such as side-channel analysis (a), fault injection attacks (b) or classical reverse engineering (c) pose a massive threat to any cryptographic implementation. Since FPGAs provide a particularly efficient platform for cryptographic hardware implementations, countermeasures against these kinds of attacks have to be investigated in order to secure FPGA-based cryptographic implementations.

\section{Contributions}   
 
The overall goal of my current research is  the creation of \emph{adaptive digital systems} with improved efficiency, reliability, and security properties. To reach this goal, my research has focused in the past years on methods and techniques to improve efficiency and reliability of FPGA-based systems by utilizing partial dynamic reconfiguration. Moreover, valuable preliminary work has been done in order to also increase the security of FPGA-based systems which has lead to a successful proposal for a three years project, funded by the \emph{Federal Ministry of Education and Research} (BMBF). In general, my research can be summarized into the following areas (see Figure \ref{fig:overview}):

\begin{figure}[t] \centering 
\includegraphics[width=0.6\textwidth]{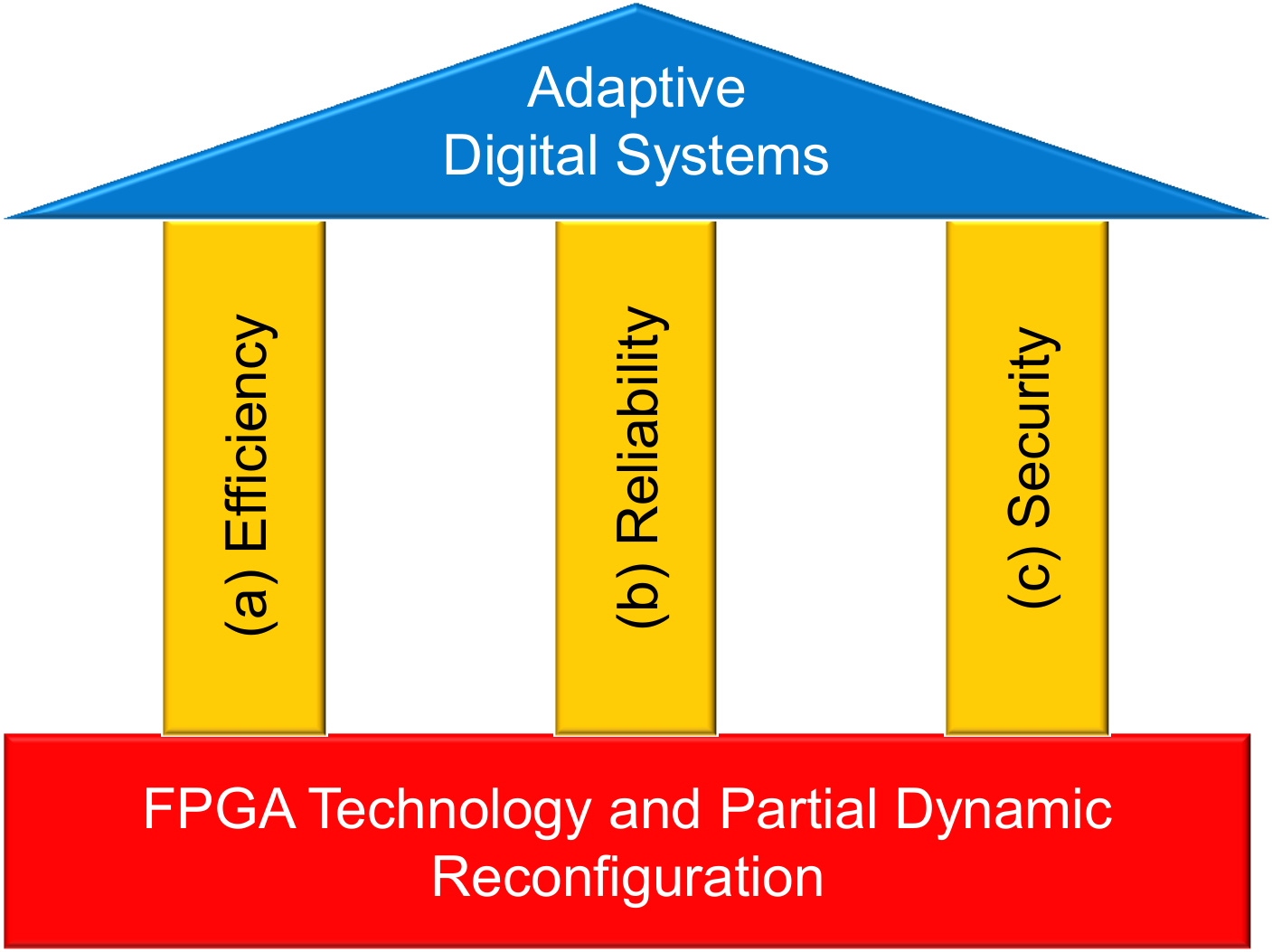} 
\caption{The building blocks of my research on adaptive digital systems: Methods to improve efficiency, reliability, and security by exploiting  the underlying FPGA technology and,  in particular, partial dynamic reconfiguration.  } \label{fig:overview}
\end{figure}

\begin{description}

\item[(a)] \textbf{Methods for Improving the Efficiency of Reconfigurable Systems:}

The flexibility and adaptivity of reconfigurable systems can be enormously enhanced by  loading different hardware modules on demand at run time. Moreover, by removing or unloading of currently not needed modules, the freed FPGA resources can be used for new tasks and, therefore, the overall resource utilization can be improved. Furthermore, the usage of pre-synthesized modules, stored in a library of \emph{partial bitstreams}, allows configuring and assembling complex data paths very quickly at run time without the need of time consuming logic synthesis and implementation. The improvements in efficiency of such adaptive systems were evaluated by different commercial example applications, like FPGA-based acceleration of SQL query processing \cite{FCCM:2012,FCCM:2013,DBLP:conf/fpl/Becher14,BZMT15, ZBB16}, image and signal processing applications  \cite{ZWOWT11,FCCM:2014}, as well as neural network accelerators \cite{PZ6}. Moreover, due to the utilization of novel techniques of FPGA-based \emph{approximate computing} \cite{EWBTZ16,BEZWT:2016,BEZT15}, the efficiency could be further improved.

\item[(b)] \textbf{Methods for Improving the Reliability of Reconfigurable Systems:}

Harsh environments for the application of FPGAs include satellite missions and avionics.   Here, especially SRAM-based FPGAs have to deal with radiation-induced errors like single event effects. One countermeasure that uses partial reconfiguration is known as \emph{scrubbing} \cite{Schmidt:2014:ANF:2554688.2554730, SZT14}, a periodic or error-triggered refreshing of the FPGA configuration from a protected configuration storage. Alternatively or in conjunction, adaptive module redundancy schemes \cite{GRBZFTH15,FCCM:2014} have been investigated. Furthermore, the possibility to reconfigure an FPGA at run time allows also interesting countermeasures against aging effects \cite{DBLP:conf/fpt/AngermeierZGT11,DBLP:conf/fpl/AngermeierZGT11}.

\item[(c)] \textbf{Improving Security by using Dynamic Hardware Reconfiguration:}

Physical attacks such as side-channel analysis, fault injection attacks, or reverse engineering  pose a massive threat to any  cryptographic implementation. Countermeasures against these attacks are the exploitation of dynamic reconfiguration. 
In this area, I have just acquired a BMBF project named \emph{Security by Reconfiguration (SecRec)}. 

\end{description}

%
%
%
%
%
%
%

The research has been carried out in collaboration with several doctoral researchers, master and bachelor students from my research group \emph{Reconfigurable Computing}. In the above mentioned research areas and projects, I have been the principal investigator and contributor of concepts. 

\section{Papers of this Treatise}

This document is a cumulative habilitation treatise. I have selected out of my 45 peer-reviewed publications, listed in Appendix \ref{sec:mypup}, the following seven papers as the key contributions of my research. The full texts of these papers are provided in Appendix~\ref{sec:mainpapers}. 

\vspace{0.5cm}

\noindent \textbf{Methods for Improving the Efficiency of Reconfigurable Systems:}

\vspace{-2mm}

\renewcommand{\arraystretch}{3.6}
\begin{tabular}{l l l} 
\textbf{FCCM'12} & \pbox{10cm}{\small Christopher Dennl, Daniel Ziener, and J{\"u}rgen Teich.  \emph{On-the-fly Composition of FPGA-Based SQL Query Accelerators Using A Partially Reconfigurable Module Library}} & \cite{FCCM:2012} \\
\textbf{FPL'14} & \pbox{10cm}{\small Andreas Becher, Florian Bauer, Daniel Ziener, and J{\"u}rgen Teich.  \emph{Energy-Aware SQL Query Acceleration through FPGA-Based Dynamic Partial Reconfiguration.}} & \cite{DBLP:conf/fpl/Becher14} \\
\textbf{FPT'15} & \pbox{10cm}{\small Andreas Becher, Daniel Ziener, Klaus Meyer-Wegener, and J{\"u}rgen Teich.   \emph{A Co-Design Approach for Accelerated SQL Query Processing via FPGA-based Data Filtering}} & \cite{BZMT15} \\
\textbf{TRETS'16} & \pbox{10cm}{\small Daniel Ziener, Florian Bauer, Andreas Becher, Christopher Dennl, Klaus Meyer-Wegener, Ute Sch{\"u}rfeld, J{\"u}rgen Teich, J{\"o}rg-Stephan Vogt, and Helmut Weber.  \emph{FPGA-Based Dynamically Reconfigurable SQL Query Processing}} & \cite{ZBB16} \\
\end{tabular}

\vspace{0.5cm}

\noindent \textbf{Methods for Improving the Reliability of Reconfigurable Systems:}
\vspace{-2mm}

\begin{tabular}{l l l} 

\textbf{FPT'11} & \pbox{10cm}{\small Josef Angermeier, Daniel Ziener, Michael Gla{\ss},  and J{\"u}rgen Teich.  \emph{Runtime Stress-aware Replica Placement on Reconfigurable Devices under Safety Constraints}} & \cite{DBLP:conf/fpt/AngermeierZGT11} \\
\textbf{FCCM'14} & \pbox{10cm}{\small Robert Glein, Bernhard Schmidt, Florian Rittner, J{\"u}rgen Teich, and Daniel Ziener.  \emph{A Self-Adaptive SEU Mitigation System for FPGAs with an Internal Block RAM Radiation Particle Sensor}} & \cite{FCCM:2014} \\
\textbf{AHS'15} & \pbox{10cm}{\small Robert Glein, Florian Rittner, Andreas Becher, Daniel Ziener, J{\"u}rgen Frickel, J{\"u}rgen Teich, and Albert Heuberger. \emph{Reliability of Space-Grade vs. COTS SRAM-Based FPGA in N-Modular Redundancy}} & \cite{GRBZFTH15} \\

\end{tabular}

\section{Structure of this Treatise}

The remainder of this document is structured as follows:

\begin{description}

\item[Chapter 2]\textbf{Designing Adaptive Reconfigurable Systems}

In this chapter, a short introduction into the generation of general partial dynamic reconfigurable systems is given.  This includes a short review of different tools and design flows. 

\item[Chapter 3]\textbf{Improving the Efficiency of Reconfigurable Systems}

\item[Chapter 4]\textbf{Improving the Reliability of Reconfigurable Systems}

In these chapters, a brief overview of the different projects and the corresponding papers is given. In each chapter, the different projects to reach either improved efficiency or reliability of reconfigurable systems are described. Each project section is logically followed by the respective paper reprints.  However, for sake of easy printing and reading, the reprints are moved to Appendix \ref{sec:mainpapers}. 

\item[Chapter 5]\textbf{Improving Security by using Dynamic Hardware Reconfiguration}

In this chapter, my ideas and concepts for improving security of cryptographic FPGA-based  implementations by utilizing partial dynamic reconfiguration is presented. Even if these ideas are not fully elaborated and no peer reviewed publications exist, they have led to a successful BMBF project proposal. 

\item[Chapter 6]\textbf{Conclusions \& Future Work}

The key contributions are summarized, future directions are identified and conclusions are provided in this chapter.

\item[Appendix A]\textbf{Bibliography}

\item[Appendix B]\textbf{Paper Reprints}

In the appendix, the general and personal bibliography and the paper reprints are provided. The personal  bibliography includes also a complete list of own papers. 

\end{description}

\chapter{Designing Adaptive Reconfigurable Systems} \label{sec:reconSys}

This chapter presents an overview of different design flows for building FPGA-based adaptive systems utilizing partial dynamic reconfiguration. First an introduction to partial dynamic reconfiguration is given, followed by different design flows. Moreover, architectural limitations of current FPGAs are listed which hinder the further increase of dynamics in such systems.

\section{Partial Dynamic Reconfiguration of FPGAs}

\emph{Dynamic reconfiguration} of FPGAs means the exchange of the FPGA configuration during runtime. \emph{Partial dynamic reconfiguration} means that parts of the configuration can be exchanged during runtime, whereas the remainder of the configuration stays active. Dynamic and especially partial reconfiguration needs additional hardware support of the configuration manager of the FPGA.

The potential to partially  reconfigure FPGAs at runtime was introduced first with the Xilinx XC6200 series \cite{ludwig1996design} around 20 years ago. Since then, Xilinx provides partial runtime reconfiguration for  all high-end FPGA series, like the Virtex series, and also later for cost-efficient FPGAs, like the Spartan family and its predecessors,  the Artix and Kintex families. Altera introduced the partial reconfiguration feature for the Stratix-5 devices in 2010 \cite{talkraw14}. Today, almost all available SRAM-based FPGAs support partial dynamic reconfiguration.   Although partial dynamic reconfiguration is widely evaluated, compared to static designs, and used by research groups since more than 20 years \cite{platzner2010dynamically, koch2012partial}, the way into industrial applications was blocked by the lack of design tools. Despite the fact that Xilinx and Altera now provide commercial design tools for partial dynamic reconfigurable systems, they are only rarely used in industrial applications.  

There exists no official statistics over the distribution of partial dynamic reconfigurable systems. However, the largest  application area of partial reconfigurable system seems to be military communication systems, with applications such as software defined radio.  In the following, some industrial applications utilizing  partial reconfiguration are listed:

According to Mike Hutton from Altera \cite{mikehuttenfpl14talk}, the reason why Altera introduced partial reconfiguration in 2010 was the need of reconfiguring client protocols of \emph{optical transport networks} (OTNs) as well as scrubbing as a mitigation against \emph{single event upsets} (SEUs) in the configuration memory. Beside these two application areas, Altera mentions also the OpenCL kernel acceleration and secure applications \cite{talkraw14}. 

Telecommunication network providers multiplex different protocols, like 10 Gb/s Ethernet or OTN-2, for different connected clients  over a faster communication network, i.e., OTN-4 with 100 Gb/s. With the help of partial reconfiguration, the client protocol stack can be dynamically exchanged while not requiring more expensive hardware or complete replacement \cite{talkraw14}.   

One problem of implementing PCIe on FPGAs is the requirement of  a fast PCIe device registration on power up. Today, the configuration files of FPGAs are so large, that it takes too long to configure the whole FPGA at once. One possible solution is the partial reconfiguration of the PCIe core followed by the PCIe registration on the host. Afterwards, the rest of the configuration may be loaded.

The conclusion is that partial reconfiguration is physically supported in FPGAs since many years. However, the usage for industrial designs is still in its infancy and not yet exploring the great opportunities which might be offered by partial dynamic reconfiguration. 

\section{Design Flows for Building Partial Reconfigurable Systems}

FPGA support for partial reconfiguration is the precondition for utilizing partial reconfiguration. However, a corresponding design flow in order to build such a system is also needed. A partial reconfigurable design is usually split into two parts: The a) \emph{static part} is always present and only configured at power up of the system. In this part, usually the interfaces to peripheral devices, memory controllers, and the access to the configuration interface of the FPGA (e.g., the ICAP for Xilinx FPGAs) is included. The configuration of one or several b) \emph{partial reconfigurable parts} or \emph{areas} can be exchanged during runtime. These areas are usually embedded and surrounded by the \emph{static part}. In these \emph{partial reconfigurable areas},  modules and operations are implemented which can be adapted or exchanged during runtime.

\begin{figure}[t] \centering 
\includegraphics[width=\textwidth]{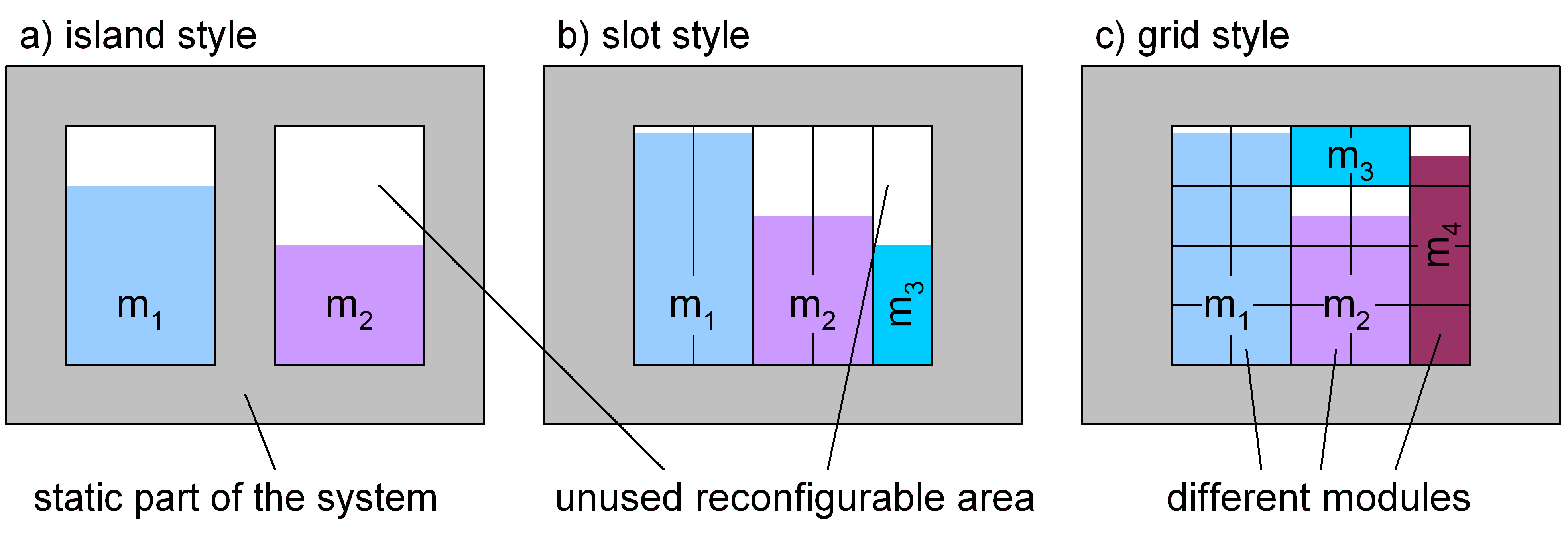} 
\caption{The different configuration styles for designing partial reconfigurable systems. On the a) \emph{island style}, only one module can exclusively be loaded  in one area at the same time. The b) \emph{slot} and c) \emph{grid style} can host multiple modules with different shapes. Taken from \cite{koch2012partial}.  } \label{fig:pstyle}
\end{figure}

The partial reconfigurable areas can be arranged in different configuration styles (see Figure \ref{fig:pstyle}). The simplest configuration style is the \emph{island style} which is capable to host one module exclusively per partial reconfigurable area. One drawback is the fragmentation if partial reconfigurable modules with different logic and routing utilization are used. The size of the partial reconfigurable area must be large enough to host all instances of the largest module which might result in a low utilization of the smaller modules. The negative effect of fragmentation can be reduced if the \emph{slot} or \emph{grid style} is used (see b) and c) in Figure \ref{fig:pstyle}). Here, the partial reconfigurable area is partitioned into slots or fields. The partial reconfigurable modules can utilize multiple slots or fields depending on the required amount of resources.

\emph{Relocation} of partial reconfigurable modules means, that the same partial configuration can be loaded on different locations onto the FPGA which makes also possible to instantiate one partial reconfigurable module multiple times on the FPGA. A very flexible hardware system can be designed by combining relocation with the slot or grid configuration style. However, such a system needs sophisticated communication structures to establish the transfer of data in and out of the partial reconfigurable area and between the different reconfigurable modules. 

Using routing resources for nets or even placement of instances belonging to the static part of the design inside partial reconfigurable areas is also possible. The advantage is the easy integration into the design flow due to relaxed routing constraints. If static instances are included in the reconfigurable module, relocation is not possible due to the fact that static instances have to be included on the same location in the configuration of all reconfigurable modules. More about different configuration styles and the corresponding communication structures as well as their realization combined with some  applications can be found in \cite{koch2012partial}. 

Xilinx and Altera offer design tools for partial reconfigurable systems and sell licenses to enable this feature. Xilinx integrated the partial reconfiguration feature in their design tool \emph{PlanAhead} \cite{planahead} and \emph{Vivado}. 
Also, Altera supports partial reconfiguration for the new \emph{Stratix V} series and integrated a partial design flow in their tools which is quite similar to the Xilinx approach \cite{bou11}. However, these approaches support only an island reconfiguration style with the inclusion of static nets in the reconfigurable areas which forbids the relocation of partial reconfigurable modules.

To overcome these restrictions, the FPGA research community has introduced some partial design flows which are able to support the more advanced slot style and module relocation.  A  very comfortable flow for building partial reconfigurable systems is the tool \emph{ReCoBus-Builder} \cite{recobus}. This tool provides the easy generation of communication structures for bus-based and data-flow-oriented communications for the slot configuration style. The successor of \emph{ReCoBus-Builder} is the tool \emph{GoAhead} \cite{goahead} which supports also newest Xilinx FPGA generations.

The approaches proposed in this treatise relies on the one hand on the tool \emph{GoAhead} \cite{goahead} for slot style reconfigurable areas for database acceleration as well as video and signal processing \cite{FCCM:2012,FCCM:2013,DBLP:conf/fpl/Becher14, ZBB16,ZWOWT11,FCCM:2014}. On the other hand, for the latest high throughput  database acceleration \cite{BZMT15}, the island style using the tool  \emph{PlanAhead} \cite{planahead} was used due to the relaxed routing constraints which corresponds in an increased maximum clock frequency. We used also a combination of both tools in \cite{ZBB16,DBLP:conf/fpl/Becher14}.

One important aspect for dynamic hardware systems is the saving of the current state or context of a partial reconfigurable module before the preemption and the state restoring during the reactivation of such a module. Due to the fact that in this treatise only stream-based reconfigurable modules are used, state saving or restoring actions are not needed which saves valuable reconfiguration time. Due to the decreased module switching time, the flexibility of such systems for high performance applications is increased.

These partial reconfigurable systems and the corresponding design flows allow implementing very dynamic, complex, and flexible systems. On such a system, different applications could be executed, where each application consists of one or more partial reconfigurable modules as well as the corresponding software running on an embedded CPU in the static system. The applications can be exchanged on such a  \emph{multi-mode system} by external triggers. If the different modes (combination of partial reconfigurable modules and corresponding software) are known at design time, a design space exploitation could be used to determine the best locations for the placements of partial reconfigurable modules and the corresponding communication structures \cite{WRZT:2013, DBLP:conf/codes/WildermannRZT11, DBLP:conf/fpl/WildermannTZ11, GYRT2012}.  

Furthermore, high-level languages, like C or C++, could be used to ease the process for developing and implementing partial reconfigurable modules. The corresponding generating of RTL descriptions is done by using \emph{High-Level Synthesis} (HLS) tools \cite{KHZ16,HKZ15,HKZ14}. The combination of HLS tools and partial reconfiguration could simplify and accelerate the development cycle for new modules in such an adaptive system.

\section{Architectural Limitations of Current FPGAs}

The current support of partial reconfiguration has also some limitations. The smallest reconfigurable part of a configuration is one \emph{frame} which is for current Xilinx device the height of one clock region (40 CLBs on Xilinx Virtex-6) and for Altera devices the height of the whole FPGA. The width of a frame is one bit. Furthermore, there is no random access to the configuration memory. Instead, the access to the configuration memory is handled over internal or external configuration ports by applying a vendor specific configuration protocol. Currently, the access to the configuration memory is limited to a single internal configuration port called \emph{internal configuration access port} (ICAP) for Xilinx FPGAs. This limitation might be circumvented by splitting a huge FPGA into several smaller ones, each with it's own internal configuration interface. By doing so, multiple partial regions may be reconfigured at the same time under the constraint that only one region is reconfigured per FPGA.   Furthermore, new FPGA architectures, like Tabula FPGAs \cite{tabula}, might overcome these limitations and offer a more efficient usage of partial reconfiguration. 

\chapter{Improving the Efficiency of Reconfigurable Systems} \label{sec:efficiency}

The efficiency of reconfigurable systems can be significantly improved by using dynamic partial reconfiguration offered by many up-to-date FPGA devices. The flexibility and adaptivity of reconfigurable systems can be enormously enhanced by loading different hardware modules on demand at run time. Moreover, by removing or unloading of currently not needed modules, the freed FPGA resources can be used for new tasks and, therefore, the overall resource utilization can be improved. Furthermore, the usage of pre-synthesized modules, stored in a library of partial bitstreams, allows configuring and assembling complex data paths very quickly at run time without the need of time consuming logic synthesis and implementation. The improvements in efficiency of such adaptive systems were evaluated by different commercial example applications, like FPGA-based acceleration of SQL query processing \cite{FCCM:2012,FCCM:2013,DBLP:conf/fpl/Becher14,BZMT15, ZBB16} as well as image and signal processing applications \cite{ZWOWT11,FCCM:2014}.

\section{FPGA-based Acceleration of SQL Query Processing}

An FPGA-based SQL query processing approach exploiting the capabilities of partial dynamic reconfiguration is presented in this section.  After the analysis of an incoming query, a query-specific hardware processing unit is generated on-the-fly and loaded on the FPGA for immediate query execution. For each query, a specialized hardware accelerator pipeline is composed and configured on the FPGA from a set of presynthesized hardware modules. These partially reconfigurable hardware modules are gathered in a library covering all major SQL operations like  restrictions, aggregations, as well as more complex operations such as joins and sorts. Moreover, this holistic query processing approach in hardware supports different data processing strategies including as well row- as column-wise data processing, in order to optimize data communication and processing. Most of the presented work in this section was done in  a   three year lasting research project together with and funded from IBM.

\subsection{Goals}

The primary goal of this project was to increase the energy efficiency and throughput of processing database queries by using adaptive FPGA-based accelerators. These goals have been reached by combining the efficiency of hardware-based accelerators with the flexibility of software-defined solutions. The flexibility has been achieved by utilizing partial dynamic reconfiguration of FPGAs.  

\subsection{Approach} \label{sec:sqlapp}

The query acceleration approach consists of a static hardware part, including communication and configuration interfaces, a library of partially reconfigurable modules which covers almost all common SQL operators, as well as software running on the host system to analyze an incoming query, select partially reconfigurable modules from the library and determine feasible placements for them, as well as controlling the communication to and configuration of the FPGA. In details, operator modules from the module library,
which reside in the main memory of the host,
are selected
and the query data path is composed on-the-fly.
After that,
the streaming data path,
which typically cascades several modules,
is loaded into a partially reconfigurable area inside the FPGA.
Furthermore, the reconfiguration manager keeps track
of the allocation of partially reconfigurable areas as well as active and finished queries. After the loading of the modules,
the database tables are streamed
from the main memory to the FPGA and into the partially reconfigurable area.
Hereby, the data is processed by the loaded operator modules
and the corresponding result is streamed continuously back to the main memory.
Each partially reconfigurable area may implement one or a subset of a query accelerator, and each may consist of one or more partially reconfigurable modules.

\begin{figure}[ht]
  \centering
  \includegraphics[width=400pt]{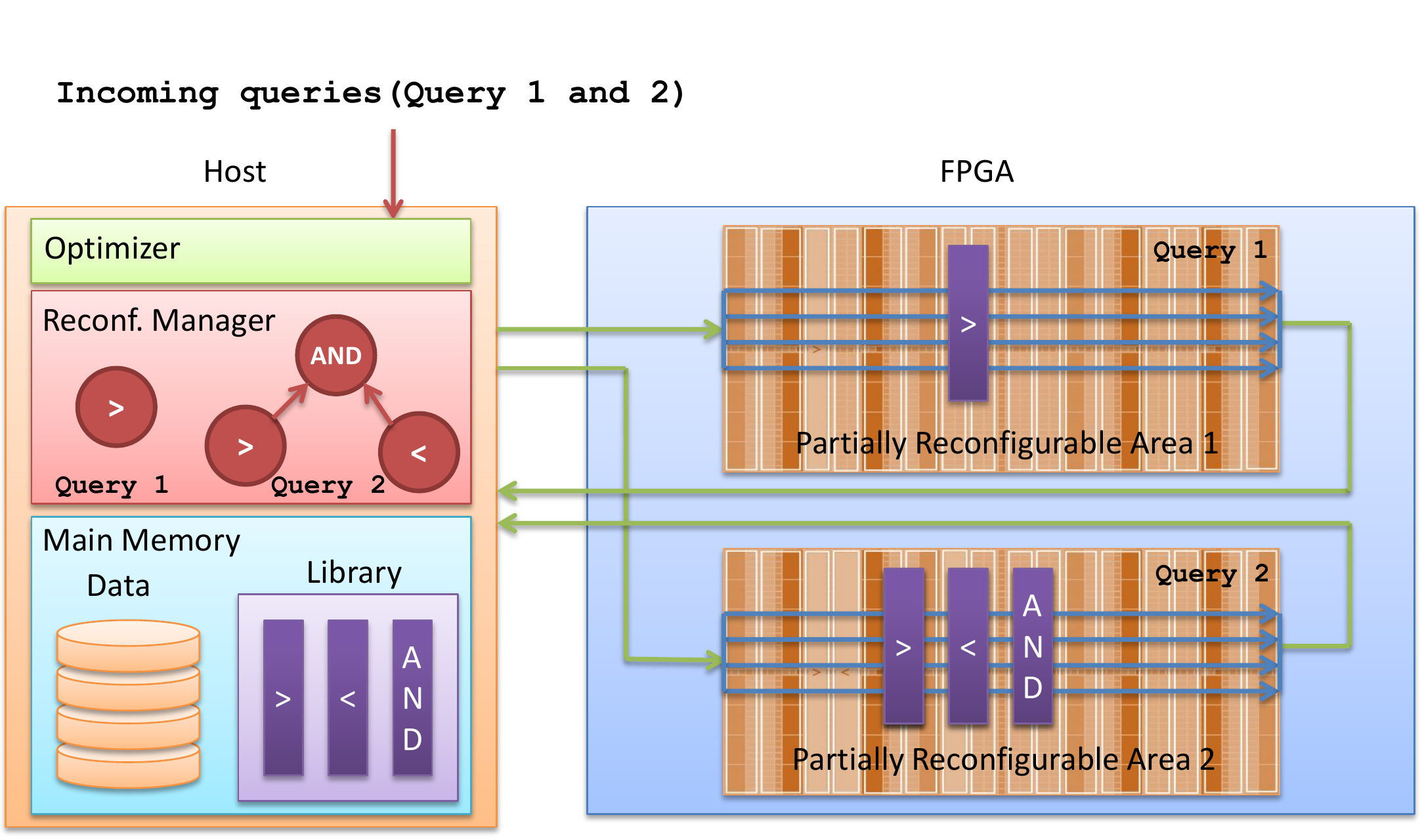}
  \caption{An overview of the  query acceleration system:
    Each incoming query is analyzed by the Reconfiguration Manager
    and corresponding modules of the generated query execution plan are loaded subsequently onto the FPGA.
    On the right,
    two partially reconfigurable areas are depicted
    which even allow the simultaneous processing of two (partial) queries in parallel. Taken from \cite{ZBB16}.}
\label{fig:overview4}
\end{figure}

Figure~\ref{fig:overview4} shows also
the partitioning of partially reconfigurable areas into elementary units called \emph{slots} 
forming a grid for the placement of modules.
Each module occupies one or more neighboring slots.
Several concatenated modules finally make a specific query accelerator.

This query processing system has evolved from a first publication of basic concepts \cite{FCCM:2012} to a mature and complete design flow for query processing as presented in \cite{ZBB16}.   Starting with just restrictions \cite{FCCM:2012}, aggregations \cite{FCCM:2013}, as well as join and sort \cite{DBLP:conf/fpl/Becher14}, the hash-join and column-based processing modules, introduced in \cite{ZBB16}, completed the portfolio of supported operations. 
Furthermore, investigations to process table data column-wise by introducing special modules  in order to reduce the amount of incoming data were analyzed. 
Moreover, a calculus for performance assessment was presented in \cite{ZBB16} which gives the possibility to evaluate the processing time of a query on different, even non-existent architectures with different parameters, e.g., throughput of communication interfaces. It also allows to explore the rich facets of different processing possibilities by combining different modules and different implementations of the same query operations, e.g., processing a join operation as a hash or rather a sort-merge-join. With the help of this performance calculus, 
we are able to chose at run time the best configuration of our SQL processing system for a given incoming query, database size, and an estimated selectivity of the query. 


The weakness of this architecture is the I/O bottleneck. Therefore, a new architecture was developed which circumvent this bottleneck \cite{BZMT15}. 
The new architecture is based on an intelligent hardware/software co-design and consists of a highly configurable FPGA-based filter chain with arithmetic operation support and an alignment unit. It feeds the filtered data directly and in a cache-optimized way to an embedded processor which is responsible for joining tables and post processing.
High throughput interfaces and parallelism of FPGAs were thus combined in order to provide reduced and cache-aligned data for optimized processor access.
As a key component, a new highly configurable bloom filter cascade was introduced to relieve a processor of time-consuming hash-value computation and to significantly reduce the data for hash joins. 

\begin{figure}[h]
\centering
\includegraphics[width=350pt]{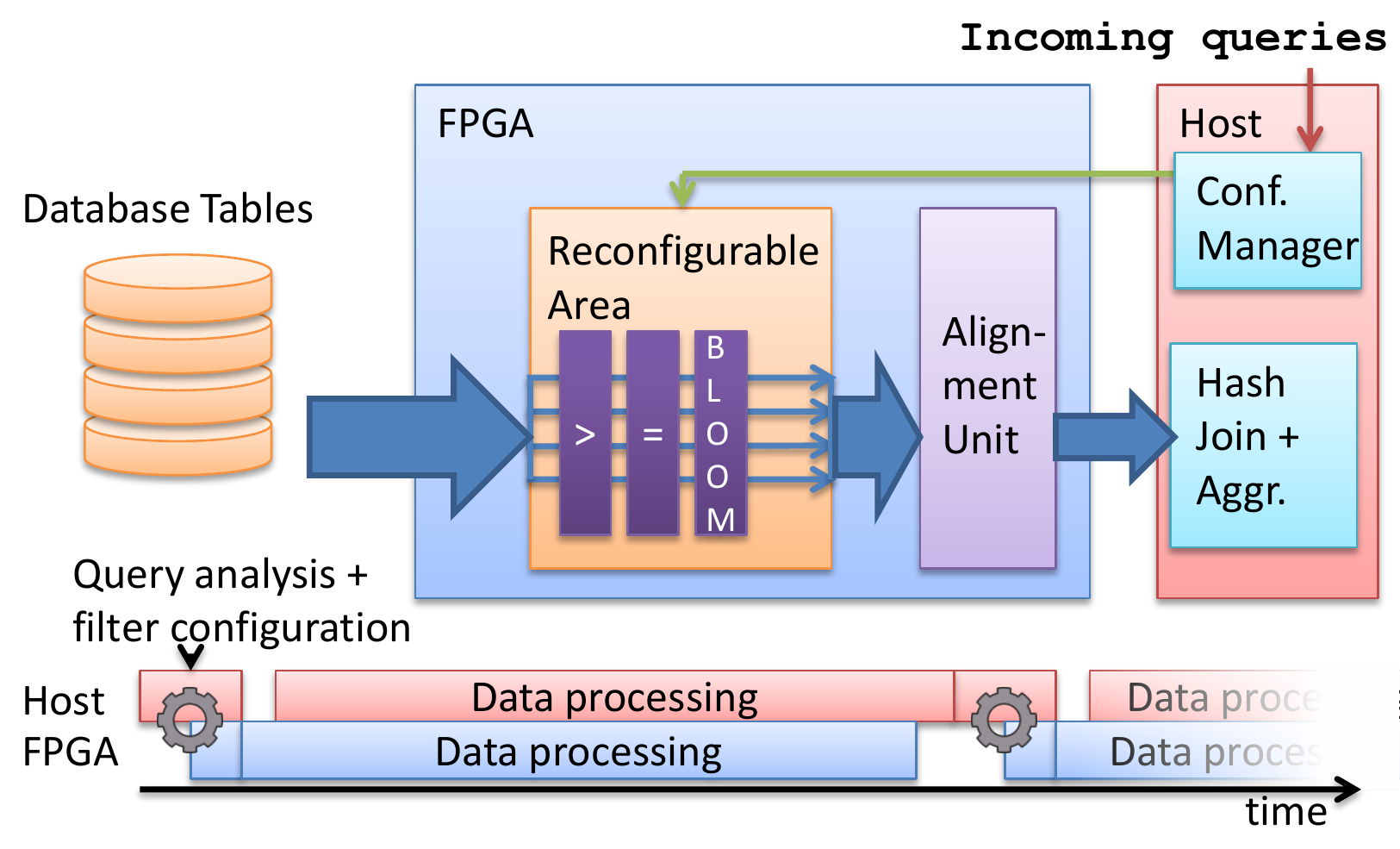}
\caption{
Overview of the new architecture for hardware-accelerated query processing (above).
First, an incoming query is analyzed and a filter chain configured by parameter adaptation and, if necessary, by structure adaptation through partial dynamic reconfiguration.
Afterwards, the data is streamed through the filter chain and in the meanwhile, the already filtered data is processed by the software implemented hash join (see timing diagram below). Taken from \cite{BZMT15}. 
}
\label{fig:architecture}
\end{figure}

Figure \ref{fig:architecture} shows an overview of the proposed co-design.
The data is fetched from external memory
by either a high-speed memory controller,
or from an SSD array via SATA connections,
and is streamed through the FPGA-based filter chain on the FPGA.
The filter chain may contain three types of modules:
a \emph{restriction} module which covers \texttt{where} clauses of a query,
an \emph{ALU} module which covers arithmetic expressions in a \texttt{where} clause,
a \emph{bloom filter} module which is responsible for pre-filtering, and hash value calculation
of the data for a subsequent hash-based join.
After the data reduction achieved by these modules, an alignment unit
adjusts the filtered data for best possible subsequent processor access, e.g., to be memory-aligned, cache-line-aligned, and cache-optimized.
The aligned remaining data is then processed by the processor system (Host), e.g., by utilizing cache coherent processor interfaces.

\subsection{Results}

By implementing the holistic query processing system presented in \cite{ZBB16},  we are able to process queries with different kind of operators,  namely restrictions, aggregations, reorder, join, and sorting. Moreover, for many operations, we implemented different algorithms to provide processing alternatives, like processing the join as hash join or as merge join. The best implementation alternative, depending on the query and data to process, could be chosen at run time which increases the flexibility enormously.  

The comparison of the measured and analyzed throughput with x86-based servers showed that the achieved throughput of our system is only comparable with processor-based variants.  The main weakness of our system lies on the limited I/O interface, in our case the PCIe or the AXI interface. However, the throughput of the internal partial reconfigurable operator pipeline is quite high. On the other hand, we need only 5\% of the energy compared to an x86-based software solution \cite{DBLP:conf/fpl/Becher14}.  Moreover, the overhead of the partial reconfiguration process was analyzed and the impact is rather low, if no exhaustive data slice scheduling is used \cite{ZBB16}. 

 To increase the throughput, we developed the new architecture which uses the FPGA part only for streaming-based operators and the more control-flow-like  hash join is processed in software by an embedded processor on an SoC device (see Section \ref{sec:sqlapp}). With this novel architecture, we outperform a x86-based system by the factor of 10 and reached a 30 times better energy efficiency \cite{BZMT15}.

\subsection{Key Papers} 

In the following, I briefly classify the role of the four related key papers for the topic: \emph{FPGA-based Acceleration of SQL Query Processing}, which are
part of this cumulative habilitation treatise. Reprints of these papers are available in
Appendix~\ref{sec:mainpapers}.

\renewcommand{\arraystretch}{3.6}
\begin{tabular}{l l} 
\pbox{4cm}{\textbf{FCCM'12}} & \pbox{11cm}{\small Christopher Dennl, Daniel Ziener, and J{\"u}rgen Teich.  \emph{On-the-fly Composition of FPGA-Based SQL Query Accelerators Using A Partially Reconfigurable Module Library} \cite{FCCM:2012}} \\
 &  \pbox{11cm}{ \vspace{5mm} This paper is the first of our papers of hardware-based SQL query  acceleration. In this paper, we introduced the technique of our partial reconfigurable areas and the mapping from a given query plan into a data path which can be loaded into these partial reconfigurable areas. We introduced modules which are able to process the SQL operators for arithmetic and restrictions. Moreover, we showed that for arithmetic intensive queries, we are faster than software solutions.  \\
 My personal contribution to this work was beside developing the ideas and concept, the (co-)supervision of the corresponding Master's Thesis, as well as writing around 40\% of the publication. } \\
 \end{tabular}
 
\renewcommand{\arraystretch}{4.6}

 \begin{tabular}{l l} 

\pbox{4cm}{\textbf{FPL'14}} & \pbox{12cm}{\small Andreas Becher, Florian Bauer, Daniel Ziener, and J{\"u}rgen Teich.  \emph{Energy-Aware SQL Query Acceleration through FPGA-Based Dynamic Partial Reconfiguration.} \cite{DBLP:conf/fpl/Becher14}} \\

 &  \pbox{12cm}{ \vspace{3mm} This paper is the third paper for our hardware-based SQL query accelerator. The focus of this paper lies on the energy efficiency of our FPGA-based accelerator. In this paper, we ported our acceleration system on the Xilinx Zynq device with an embedded ARM processor. We further introduced in this paper the more complex operations join and sort. Furthermore, a reordering module was introduced which is able to reorder, insert, or remove attributes within a tuple. Next, a mathematical throughput analysis was presented for performance analysis of a query plan based on the provided query module library.\\
 My personal contribution to this work was beside developing the basic concept, the (co-)supervision of the two corresponding Master's Theses, the verification of the results, as well as writing around 60\% of the publication. } \\


\pbox{4cm}{\textbf{FPT'15}} & \pbox{12cm}{\small Andreas Becher, Daniel Ziener, Klaus Meyer-Wegener, and J{\"u}rgen Teich.   \emph{A Co-Design Approach for Accelerated SQL Query Processing via FPGA-based Data Filtering} \cite{BZMT15} }\\
 &  \pbox{12cm}{ \vspace{3mm}  In this paper, we presented the new architecture based on an intelligent hardware/software co-design in order to speed up the data processing compared to our first approach. The FPGA-based hardware is able to filter and preprocess data at full memory throughput. The amount of resulting data is extremely reduced by the hardware filters.  Therefore, it can be easily further processed by an embedded processor without introducing a bottleneck.  The processor is responsible for  control-intensive part, like join processing. Our embedded implementation is up to ten times faster than an implementation on a full-featured x86 processor.  \\
  My personal contribution to this work was beside developing the basic concept, the (co-)supervision of the PhD student, as well as writing extensive parts of the publication. 
 } \\

  \end{tabular}
 
\renewcommand{\arraystretch}{4.6}

 \begin{tabular}{l l} 
 

 
\pbox{4cm}{\textbf{TRETS'16}} & \pbox{12cm}{\small Daniel Ziener, Florian Bauer, Andreas Becher, Christopher Dennl, Klaus Meyer-Wegener, Ute Sch{\"u}rfeld, J{\"u}rgen Teich, J{\"o}rg-Stephan Vogt, and Helmut Weber.  \emph{FPGA-Based Dynamically Reconfigurable SQL Query Processing} \cite{ZBB16}} \\
 &  \pbox{12cm}{ \vspace{3mm}  This journal paper gives a good overview about our work for energy-efficient accelerators for SQL query processing. This paper summarizes the  achievements of  a three year project funded by IBM. Therefore, also our partners from IBM are on the authors list. In this paper, we presented our acceleration system and the investigated SQL operators and their corresponding implementations. The new contributions above the former papers are the hash join, the complete calculus for the performance estimation and the overhead introduced by the partial reconfiguration. Note that the system described in  paper above \cite{BZMT15} is finalized after the end of the project and, therefore, not included in this publication.    \\
  My personal contribution to this work was the structure and content of the publication, the arrangement and revision of different text paragraphs as well as the writing of extensive parts of the publication. } \\

\end{tabular}

\subsection{Team \& Supervised Theses}

Publications \cite{FCCM:2012} and \cite{FCCM:2013} was conducted with Christopher Dennl, a former PhD student in my group. Publications \cite{DBLP:conf/fpl/Becher14} and \cite{BZMT15} was conducted with Andreas Becher, also a PhD student in my group.  In this research area, I (co-)supervised the following theses:

\begin{itemize}

\item Christopher Dennl, Diplomarbeit, \emph{Aufbau einer SQL-Operator-Bibliothek bestehend aus partiell rekonfigurierbaren Modulen} (engl. Development of an SQL Operator Library Consisting of Partially Reconfigurable Modules), Hardware/Software Co-Design, FAU Er\-lang\-en-N{\"u}rnberg, November 2011

\item Anna Sch{\"u}pferling, Studienarbeit, \emph{Automatische Makroerzeugung f{\"u}r die dynamisch partielle Rekonfiguration von FPGAs} (engl. Automatic Macro Generation for FPGA-based Dynamic Partial Reconfiguration), Hardware/Software Co-Design, FAU Er\-lang\-en-N{\"u}rnberg, June 2012

\item Christian Knell, Bachelorarbeit,	\emph{Modellrechungen f{\"u}r die Ausf{\"u}hrung von Bus\-i\-ness-Analytic-Anfragen mit Hilfe eines dynamisch
rekonfigurierbaren FPGAs}  (engl. Performance Estimation for Processing of Business Analytic Queries on Dynamic
Reconfigurable FPGAs), Data Management, FAU Erlangen-N{\"u}rnberg, April 2013

\item Florian Bauer, Projektarbeit, \emph{Entwurf eines dynamisch partiell rekonfigurierbaren Datenbankbeschleunigers} (engl. Design of a Dynamically Partially Reconfigurable Database Accelerator), Hardware/Software Co-Design, FAU Er\-lang\-en-N{\"u}rnberg, March 2014

\item Florian Bauer, Masterarbeit, \emph{Concepts and Implementation of an FPGA-based SQL Accelerator for Processing Column Store Tables}, Hardware/Software Co-Design, FAU Erlangen-N{\"u}rnberg, June 2014

\item Micha Schie{\ss}l, Bachelorarbeit, \emph{Rekonfigurationsmanager f{\"u}r partiell rekonfigurierbare Datenbankbeschleuniger} (engl. Reconfiguration Manager for Partial Reconfigurable Database Accelerator), Hardware/Software Co-Design, FAU Er\-lang\-en-N{\"u}rnberg, July 2014

\item Andreas Becher, Projektarbeit,	\emph{Beschleunigung von SQL-Joins auf FPGAs} (engl. Acceleration of SQL Joins on FPGAs), Hardware/Software Co-Design, FAU Erlangen-N{\"u}rnberg, September 2014

\item Andreas Becher, Masterarbeit, \emph{FPGA-based Implementation of Energy Efficient Hash Join Operations for SQL Queries}, Hardware/Software Co-Design, FAU Erlangen-N{\"u}rnberg, September 2014

\item Tobias Alscher, Bachelorarbeit, \emph{Effiziente Datenstrukturen zur Erm{\"o}glichung von Abfragen f{\"u}r einen Hardware-basierten Key-Value-Store} (engl. Efficient Data Structures for Queries on a Hardware-based Key-Value Store), Institute of Embedded Systems, TU Hamburg, October 2016

\end{itemize}

\section{Image and Signal Processing Applications}

Beside the FPGA-based acceleration of SQL query processing, we investigated also the  improvement of efficiency  for image and signal processing applications. These activities are summarized in this section.  

\subsection{Approaches}

In this section, image \cite{ZWOWT11,KZF12} and signal processing approaches \cite{FCCM:2014,GRBZFTH15,KZF12} are presented which goal is to improve the efficiency for reconfigurable systems. This includes also techniques as approximate computing \cite{EWBTZ16,BEZWT:2016,BEZT15} and efficient processing of neural networks in reconfiguration hardware \cite{PZ6}. 

\subsubsection{Image Processing}
An FPGA-based smart camera system with support for dynamic run-time reconfiguration was presented in \cite{ZWOWT11} and \cite{KZF12}. The underlying architecture consists of a static SoC which can be extended by dynamic modules. These modules are responsible for the stream-based image processing and can be loaded and unloaded at run-time.  Modules for detecting skin colors and image filtering are implemented as well as a frame buffer, particle filter, motion detection, and pixel marker module. Furthermore, even the position of these modules in the processing chain can be exchanged. Later, the module repository is extended by sobel filtering modules, a background classification module, and an alarm region module.

\subsubsection{Signal Processing}

An FPGA-based, dynamically reconfigurable Software Defined Radio (SDR) platform was proposed in \cite{KZF12} to enable fast multi-mode and multi-standard switching in legacy and future wireless transmission networks. The SDR signal processing chain, consisting of different partial modules from a hardware library, can be easily adapted at run-time in order to exchange communication standards or parameters by using partial reconfiguration with dedicated communication structures. The library consists of general purpose SDR and application-specific modules. The introduced general purpose modules support beside structural reconfiguration also fast behavioral adaptation by changing parameters over a bus interface.  Using this module library, a huge variety of DSP systems can be realized very fast without the need of module re-design and re-synthesis. This allows the ideal exertion as a rapid-prototyping platform for DSP applications. Furthermore, by using parameter adaptation of the partial modules, the adaption time can be further lowered which allows the deployment in adaptive communication systems with support of multiple communication standards, e.g., baseband transmitters.

In order to mitigate Single Event Upsets in the FPGA configuration and fabric and to have a very flexible communication system,  a self-adaptive FPGA-based, partially reconfigurable system for space missions was presented in \cite{FCCM:2014} and \cite{GRBZFTH15}. Dynamic reconfiguration is used here for an on-demand replication of modules in dependence of current and changing radiation levels and to update communication protocols over the envisaged life time of the satellite mission. The main focus of these approaches lies on reliability improvements. Therefore, more about this application can be found in Chapter \ref{sec:relia}. 

\subsubsection{Approximate Computing}

As a first step towards efficiency increasing by approximate computing, approximate adder structures for FPGA-based implementations were proposed in \cite{BEZT15}, \cite{BEZWT:2016}, and \cite{EWBTZ16}. These adder structures take advantage of the available FPGA resources and can significantly increase the efficiency, if the application can tolerate some deviations in the results.  Compared with a full featured accurate adder, the longest path is significantly shortened which enables the clocking with an increased clock frequency. By using the proposed adder structures, the throughput of an FPGA-based implementation can be significantly increased. On the other hand, the resulting average error can be reduced compared to similar approaches for ASIC implementations.

\subsubsection{Deep Neural Network Acceleration}

Deep neural networks are an extremely successful and widely used technique for various pattern recognition and machine learning tasks. Due to power and resource constraints, these computationally intensive networks are difficult to implement in embedded systems. Yet, the number of applications that can benefit from the mentioned possibilities is rapidly rising. A novel architecture for processing previously learned and arbitrary deep neural networks on FPGA-based SoCs was proposed in \cite{PZ6} that is able to overcome these limitations. A key contribution of our approach, which we refer to as batch processing, achieved a mitigation of required weight matrix transfers from external memory by reusing weights across multiple input samples. This technique combined with a sophisticated pipelining and the usage of high performance interfaces accelerates the data processing compared to existing approaches on the same FPGA device by one order of magnitude. Furthermore, we achieved a comparable data throughput as a fully featured x86-based system at only a fraction of its energy consumption.

\subsection{Team \& Supervised Theses}

Publications \cite{FCCM:2014} and \cite{KZF12} was conducted with Bernhard Schmidt, a former PhD student in my group. The publication \cite{GRBZFTH15} was conducted with Andreas Becher, also a PhD student in my group and Robert Glein and Florian Rittner from the Chair of  Information Technology (Communication Electronics) of the Friedrich-Alexander-Universit\"at Erlangen-N\"urnberg (FAU).   Publications  \cite{BEZT15}, \cite{BEZWT:2016}, and \cite{EWBTZ16}  was conducted with Andreas Becher and Jorge Echavarria, two PhD students in my group. Whereas, the publication  \cite{PZ6}  was conducted with Thorbj{\"o}rn Posewsky, a PhD student in my group in Hamburg. 
In this research area, I (co-)supervised the following theses:

\begin{itemize}

\item Volker Breuer, Diplomarbeit, \emph{Entwicklung eines FPGA-basierten, dynamisch 							
rekonfigurierbaren Funksystems} (engl. Development of an FPGA-based Dynamic Reconfigurable Software Defined Radio System), Hardware/Software Co-Design, FAU Er\-lang\-en-N{\"u}rnberg, February 2012

\item Christian Reinbrecht, Bachelorarbeit, \emph{Entwurf und Umsetzung von Algorithmen zur Detektion von 						sich bewegenden Objekten in FPGA-basierten Video\-{\"u}ber\-wachungs\-systemen} (engl. Design and Implementation of Algorithms for Detecting Moving Objects in FPGA-based Video Monitoring Systems), Hardware/Software Co-Design, FAU Er\-lang\-en-N{\"u}rnberg, May 2012	

\item Thomas Bartsch, Studienarbeit,  \emph{Entwurf einer FPGA-basierten Datenkonsolidierungseinheit f{\"u}r die Avionik} (engl. Design of an FPGA-based Data Consolidation Unit for Avonic Applications), Hardware/Software Co-Design, FAU Er\-lang\-en-N{\"u}rnberg, June 2012

\item Markus Blocherer, Diplomarbeit, \emph{Entwicklung einer FPGA-basierten Konsolidierungseinheit f{\"u}r 						Flie{\ss}komma- und Ganzzahlendaten im Einsatzbereich der zivilen Luftfahrt} 	(engl. Development of an FPGA-based Data Consolidation Unit for Floating Point and Integer Data for Civil Air Planes), Hardware/Software Co-Design, FAU Er\-lang\-en-N{\"u}rnberg, January 2013

\item Jutta Pirkl, Projektarbeit, \emph{Entwicklung einer Evaluationsplattform f{\"u}r FPGA-basierte
Bilderverarbeitungsmodule} (engl. Development of an Evaluation Platform for FPGA-based Image Processing Modules), Hardware/Software Co-Design, FAU Er\-lang\-en-N{\"u}rnberg, November 2015

\item Anton Heinze, Bachelorarbeit, \emph{Entwurf von partiell dynamisch rekonfigurierbaren Modulen mittels Architektursynthese} (engl. Using High-Level Synthesis for Designing Partial Dynamic Reconfigurable Modules), Institute of Embedded Systems, TU Hamburg, October 2016

\item Tobias Wessel, Masterarbeit, \emph{Optimizing Data Transfers for Deep Neural Network Implementations on SoC-FPGAs}, Institute of Embedded Systems, TU Hamburg, December 2016

\end{itemize}

\chapter{Improving the Reliability of Reconfigurable Systems} \label{sec:relia}

The success makes FPGAs also interesting for new safety-critical applications and application fields, like in space missions or avionics. However, in these new operation sites, FPGAs have to deal with harsh environments, and the implemented systems are forced to guarantee a high reliability. Especially SRAM-based FPGAs have to deal with radiation-induced errors like single event effects. Measures using the partial reconfiguration feature like scrubbing \cite{Schmidt:2014:ANF:2554688.2554730, SZT14}, a periodical or error triggered loading of the not falsified FPGA configuration, and the usage of adaptive module redundancy schemes \cite{GRBZFTH15,FCCM:2014} have been investigated. Furthermore, the possibility to reconfigure an FPGA at run time allows also interesting countermeasures for aging effects \cite{DBLP:conf/fpt/AngermeierZGT11,DBLP:conf/fpl/AngermeierZGT11}. 

\section{SEU Mitigation Techniques for FPGA-based Satellite Systems} \label{sec:relia:seu}

The improvement in reliability by using partial dynamic reconfiguration and scrubbing for SRAM-FPGA-based satellite systems has been investigated. A self-adaptive system was proposed which monitors the current SEU rate and exploits the partial reconfiguration of FPGAs to implement redundancy on demand \cite{GRBZFTH15,FCCM:2014}. The reliability can be further improved by using configuration scrubbing. Here, several enhancements were proposed which lower the scrubbing effort and increase the \emph{Mean-Time-To-Repair} (MTTR) \cite{Schmidt:2014:ANF:2554688.2554730, SZT14}.

\subsection{Goals}

The main goals of these projects were the increasing of reliability of FPGA-based satellite systems in terms of \emph{Mean-Time-To-Failure} (MTTF) or \emph{probability of failures per hour} (PFH), and the \emph{Mean-Time-To-Repair} (MTTR).  Since the intensity of cosmic rays is not constant but may vary over several magnitudes depending on the solar activity, this worst-case radiation protection is far too expensive, if redundant FPGA resources are allocated over the whole mission time. As a remedy for such inefficiency, a self-adaptive system was proposed which monitors the current SEU rate and exploits the partial reconfiguration of FPGAs to implement redundancy on demand.

\subsection{Approaches}

A self-adaptive FPGA-based, partially reconfigurable system for space missions was proposed in \cite{FCCM:2014} in order to mitigate \textit{Single Event Upsets} in the FPGA configuration and fabric. Dynamic reconfiguration is used here for an on-demand replication of modules in dependence of current and changing radiation levels. More precisely, the idea is to trigger a redundancy scheme such as \textit{Dual Modular Redundancy} or \textit{Triple Modular Redundancy} in response to a continuously monitored \textit{Single Event Upset} rate measured inside the on-chip memories itself, e.g., any subset (even used) internal \textit{Block RAMs}. Depending on the current radiation level, the minimal number of replicas is determined at run-time under the constraint that a required \textit{Safety Integrity Level} (SIL) for a module is ensured and configured accordingly. 

\begin{figure}[t]
\centering
\includegraphics[scale=0.65]{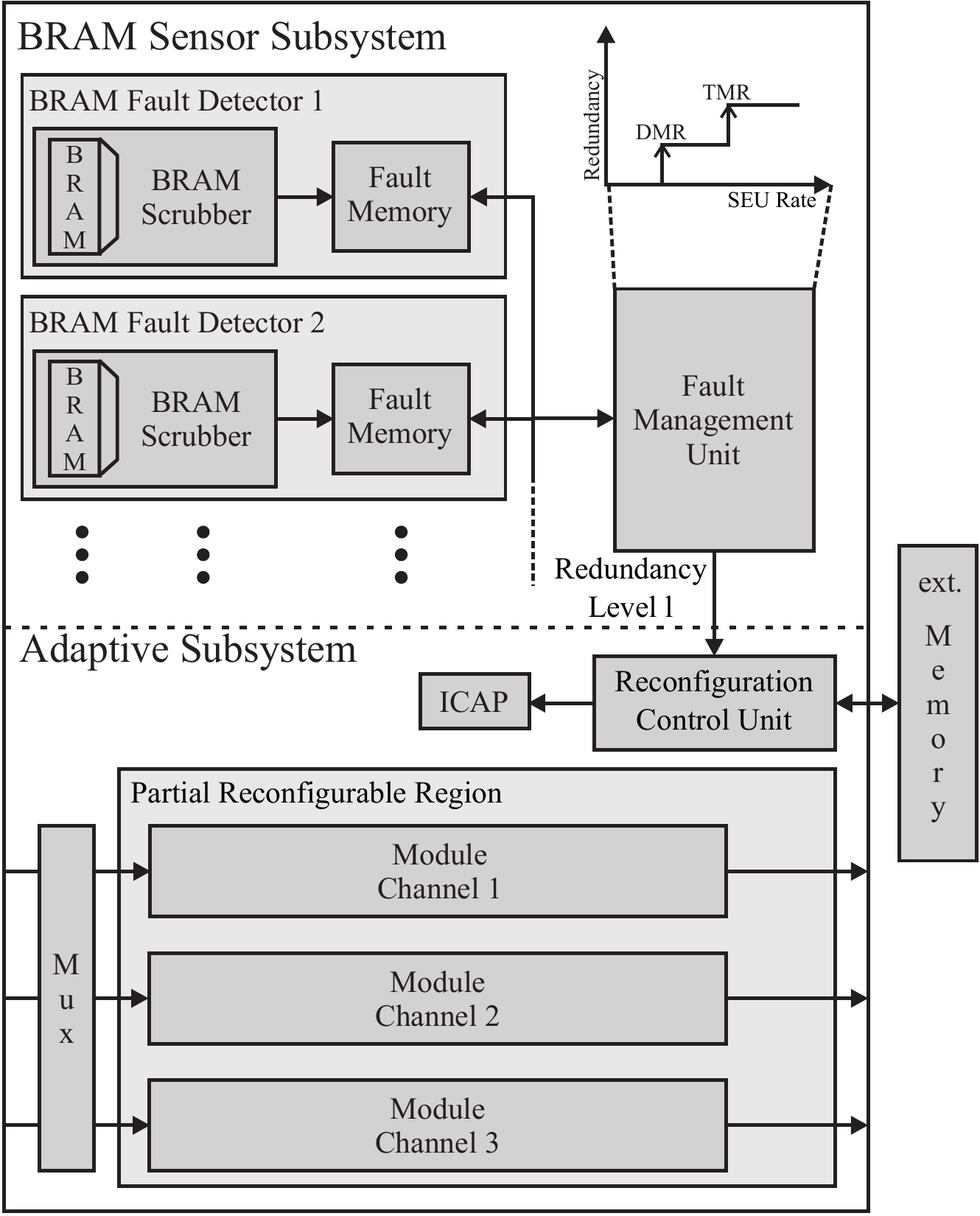}
\caption{An FPGA-based self-adaptive autonomous SEU mitigation system consisting of a \textit{BRAM Sensor Subsystem} (top) and an \textit{Adaptive Subsystem} (bottom). Taken from \cite{FCCM:2014}.
}
\label{fig:system_overview}
\end{figure}

As depicted in Fig. \ref{fig:system_overview}, our FPGA-based system consists of two subsystems: a) a \textit{BRAM Sensor Subsystem} which utilizes embedded BRAMs to estimate the current SEU rate of the configuration memory, and b) an \textit{Adaptive Subsystem} with a partially reconfigurable area which hosts the modules of the implemented application whereat the introduced redundancy level is controlled according to the current \textit{Estimated Configuration Memory SEU Rate}.

The \textit{BRAM Sensor Subsystem}, as show in Fig. \ref{fig:system_overview}, consists of a) at least one \textit{BRAM Fault Detector} (BFD) and b) a \textit{Fault Management Unit} (FMU). 
The BFD consists of a \textit{BRAM Scrubber} which continuously reads out and checks the content of one embedded BRAM block. Moreover, the \textit{BRAM Scrubber} contains an address counter to cyclically check each data word at the output port of the BRAM. Via the ECC parity bits, the \textit{BRAM Scrubber} is able to immediately correct single bit errors and detect double bit errors whereat each detected single and double bit error are accumulated separately in counters of the \textit{Fault Memory}. The counter values of each \textit{Fault Memory} are accessed by the FMU through a proprietary bus system to calculate the current SEU rate $\mu_{\text{BRAM}}$ of the embedded BRAM. On the basis of $\mu_{\text{BRAM}}$, the FMU estimates the SEU rate $\mu_{\text{CFG}}$ of the configuration memory and determines the level of redundancy as a function of $\mu_{\text{CFG}}$ and a target PFH value which might be specified by a required SIL. In general, the PFH value also depends on the size of the implemented design which is commonly measured by the number of used configuration bits.

The required level of redundancy is then signaled to the input port of the \textit{Reconfiguration Control Unit} (RCU) which belongs to the \textit{Adaptive Subsystem} and controls the number of replicas via \textit{Internal Configuration Access Port} (ICAP) by loading partial bitstreams of module replicas from an external memory into the configuration memory. In Fig. \ref{fig:system_overview}, an example \textit{Adaptive Subsystem} configuration is shown consisting of three data channels with Channel 1 having the highest priority. Channel 2 and 3 can be switched off, if the resources are needed to replicate modules of Channel 1. In the shown scenario, no replicas and, therefore, no voters are assumed in case of a low SEU rate $\mu_{\text{CFG}}$. The data of all three data channels are processed in parallel to reach the maximum achievable throughput and area utilization of the system. 

Furthermore, the suitability of different SRAM-based FPGAs for harsh radiation environments (e.g., space) was evaluated in \cite{GRBZFTH15}. In particular, we compared the space-grade and radiation-hardened by design \mbox{\textit{Virtex-5QV}} (XQR5VFX130) with the commercial off-the-shelf \mbox{\textit{Kintex-7}} (KC7K325T) from \textit{Xilinx}. The advantages of the latter device are: 2.5 times the resources of the space-grade FPGA, faster switching times, less power consumption, and the support of modern design tools. 
We focused on resource consumption as well as reliability in dependence of single event upset rates for a geostationary earth orbit satellite application, the \textit{Heinrich Hertz} satellite mission.
For this mission, we compared different modular redundancy schemes with different voter structures for the qualification of a digital communication receiver.
A major drawback of the \mbox{\textit{Kintex-7}} are current-step single event latch-ups, which are a risk for space missions. If the use of an external voter is not possible, we suggest triple modular redundancy with one single voter at the end, whereby the \mbox{\textit{Virtex-5QV}} in this configuration is about as reliable as the \mbox{\textit{Kintex-7}} in an N-modular redundancy configuration with an external high-reliable voter.

\begin{figure}[htp]
\centering
\includegraphics[width=3.2in]{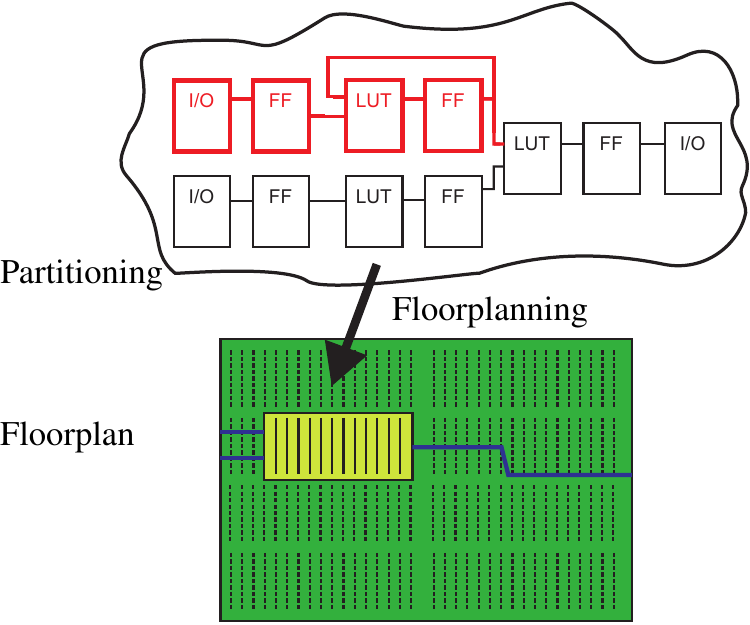}
\caption{Illustration of our two step approach. In the partitioning step, the primitive cells of the netlist, e.g., LUTs and flip-flops, and nets are categorized into \emph{essential} (black) and \emph{critical} (red) cells, nets respectively in order to identify and distinguish the associated \emph{essential} and \emph{critical} bits. In the floorplanning step, the primitive cells are placed and routed such to minimize the number of occupied configuration frames by using of special placement and routing constraints. Taken from \cite{Schmidt:2014:ANF:2554688.2554730}. }
\label{intropic}
\end{figure}

The reliability of such a system can be further increased by using configuration scrubbing \cite{hsw09, XSEM12, SariP11}. However,
existing scrubbing techniques for SEU mitigation on FPGAs do not guarantee an error-free operation after SEU recovering if the affected configuration bits do belong to feedback loops of the implemented circuits.
A netlist-based circuit analysis technique was proposed in \cite{Schmidt:2014:ANF:2554688.2554730, SZT14} to distinguish so-called \emph{critical} configuration bits from \emph{essential} bits in order to identify configuration bits which will need also state-restoring actions after a recovered SEU. 
Furthermore,  a floorplanning approach for reducing the effective number of scrubbed frames was also proposed (see Figure \ref{intropic}).

We achieved the first goal by netlist analysis with subsequent partitioning of primitive cells and nets into \emph{critical} and \emph{non-critical} cells and nets. With the help of the Xilinx tool \emph{bitgen}, we are able to determine the corresponding \emph{critical} bits in a given bitfile. The great advantage of our method over previous fault-injection approaches like \cite{lee12}, is the automatic determination of \emph{critical} bits without requiring any time-consuming bit-wise fault injection and complex verification techniques. 
Verifying and correcting bits can only be done frame-wise by reading or writing whole frames.
Therefore, the second goal, the reduction of the number of occupied frames, is achieved by manipulated floorplanning in such a way that a high frame utilization is achieved. As an important side effect, this also may lead to a reduction of the MTTR of a given system due to shorter scrubbing cycles.

\subsection{Results}

For signal processing applications it was shown that this autonomous adaption to the different solar conditions realizes a resource efficient mitigation. In our case study, we showed that it is possible to triplicate the data throughput at the \textit{Solar Maximum} condition (no flares) compared to a \textit{Triple Modular Redundancy} implementation of a single module. We also showed the decreasing \textit{Probability of Failures Per Hour} by $ 2 \times 10^{4} $ at flare-enhanced conditions compared with a non-redundant system.

The experimental results for our netlist classification and floorplanning  approach gave evidence that our optimization methodology not only allows to detect errors earlier but also to minimize the \emph{Mean-Time-To-Repair} (MTTR) of a circuit considerably.
In particular, we showed that by using our approach, the MTTR for datapath-intensive circuits can be reduced by up to 48.5 \% in comparison to standard approaches.

\subsection{Key Papers} 

In the following, I briefly classify the role of the two related key papers for the topic: \emph{SEU Mitigation Techniques for FPGA-based Satellite Systems}, which are
part of this cumulative habilitation treatise. Reprints of these papers are available in
Appendix~\ref{sec:mainpapers}.

\renewcommand{\arraystretch}{4.6}
\begin{tabular}{l l} 
 \pbox{4cm}{\textbf{FCCM'14}} & \pbox{11cm}{\small Robert Glein, Bernhard Schmidt, Florian Rittner, J{\"u}rgen Teich, and Daniel Ziener.  \emph{A Self-Adaptive SEU Mitigation System for FPGAs with an Internal Block RAM Radiation Particle Sensor} \cite{FCCM:2014}} \\
 &  \pbox{11cm}{ \vspace{5mm} This paper describes our self-adaptive FPGA-based signal processing system for  satellite applications. The current radiation is measured on-chip by utilizing BRAMs, and dependent of the measured radiation, the redundancy grade is adapted by using partial reconfiguration. This was a joint work of a Fraunhofer group which is involved into the German \emph{Heinrich Hertz Satellite} project and my group on the university. }\\
 \end{tabular}

\begin{tabular}{l l} 

\pbox{4cm}{\phantom{\textbf{FCCM'11}}} &  \pbox{11cm}{ \vspace{5mm}
 My personal contribution to this work was the substantially participation on the concept development,  the calculation of the reliability of redundant module in conjunction with scrubbing techniques, the writing of extensive parts of the publication, as well as the (co-)supervision of the  PhD student. } \\
 
\pbox{4cm}{\textbf{AHS'15}} & \pbox{11cm}{\small Robert Glein, Florian Rittner, Andreas Becher, Daniel Ziener, J{\"u}rgen Frickel, J{\"u}rgen Teich, and Albert Heuberger. \emph{Reliability of Space-Grade vs. COTS SRAM-Based FPGA in N-Modular Redundancy} \cite{GRBZFTH15}} \\
 &  \pbox{11cm}{ \vspace{5mm}   In this paper, we compared the space-grade and radiation-hardened by design \mbox{\textit{Virtex-5QV}} (XQR5VFX130) with the commercial off-the-shelf FPGA \mbox{\textit{Kintex-7}} (KC7K325T) from \textit{Xilinx}. Advantages and drawbacks of both families were discussed. We analyzed and calculated the radiation effect upset rates for the  German \emph{Heinrich Hertz Satellite} GEO satellite mission. N-modular redundancy schemes with different voter and segmentation were applied to a communication receiver FPGA design. This is the second paper from our cooperation with  the Fraunhofer group and received the \emph{Best Application Paper Award} on the AHS'15. \\
 My personal contribution to this work was the substantially participation on the concept development,  the calculation of the reliability values, the writing of extensive parts of the publication, as well as the (co-)supervision of the  PhD student. } \\

 \end{tabular}

\subsection{Team \& Supervised Theses}

The publication \cite{FCCM:2014} was conducted with Bernhard Schmidt, a former PhD student in my group and the Fraunhofer group involved in the \emph{Heinrich Hertz Satellite} project consisting of Robert Glein and Florian Rittner. The publication \cite{GRBZFTH15} was conducted with Andreas Becher, also a PhD student in my group and the same Fraunhofer group.  In this research area, I (co-)supervised the following theses:

\begin{itemize}

\item Christian Z{\"o}llner, Masterarbeit, \emph{Entwicklung und Umsetzung eines Scrubbing-Kontrollers zur Korrektur von Single-Event-Upsets im
Konfigurationsspeicher SRAM-basierter FPGAs} (engl. Development and Implementation of a Scrubbing Controller for the Correction of Single Event Upsets in Configuration memories of SRAM-based FPGAs), Hardware/Software Co-Design, FAU Er\-lang\-en-N{\"u}rnberg, December 2013

\item Michael Moese, Studienarbeit, \emph{Konzepte und Modelle zur Synchronisation von DMR ausgelegter Prozessoren}  (engl. Concepts and Models for Synchronizing DMR Configured Processors), Hardware/Software Co-Design, FAU Er\-lang\-en-N{\"u}rnberg, March 2014

\item Alexander Butiu, Masterarbeit, \emph{Entwicklung und Umsetzung eines FPGA-ba\-sier\-ten Videoverarbeitungssystems mit adaptiver Redundanzsteuerung zur Detektion und Maskierung von strahlungsinduzierten Fehlern} (engl. Development and Implementation of an FPGA-based Image Processing System with Adaptive Redundancy Control for Detection and Masking of Radiation-induced Faults), Hardware/Software Co-Design, FAU Er\-lang\-en-N{\"u}rnberg, August 2014

\item Alexander Rosenberger, Masterarbeit, \emph{Self-Adaptive SEU Mitigation for FPGAs using Partial Dynamic Reconfiguration}, Hardware/Software Co-Design, FAU Er\-lang\-en-N{\"u}rnberg, September 2015

\end{itemize}

\section{Wear-leveling for FPGA-based Systems to Mitigate Aging Effects}

In order to increase the lifetime of a reconfigurable device, we proposed in \cite{DBLP:conf/fpl/AngermeierZGT11} a placement strategy to distribute the stress equally on the reconfigurable resources at run time such that all have a similar level of degradation. Thereby, we presented a new aging model which is applied to estimate the influence of aging effects on dynamically reconfigurable devices, and which can be evaluated at run time, while providing quite accurate aging results.

\subsection{Goals}

The primary goal of this project was to increase the lifetime of reconfigurable devices by using a wear-leveling strategy for the dynamic placement of partial reconfigurable modules. The second goal was the combination of this wear-leveling strategy with a \emph{triple modular redundancy} (TMR) approach \cite{DBLP:conf/fpt/AngermeierZGT11} in order to increase the lifetime and the MTTF for radiation induces faults (see also Section \ref{sec:relia:seu}). 

\subsection{Approach}

The decreased reliability of each individual transistor with every future generation of semiconductor technology accelerates degeneration effects of these transistors. Discovered  degeneration effects in today's CMOS technologies are the \textit{hot-carrier injection effect} (HCI) \cite{gue07},  \textit{electromigration} \cite{cla90},  the \textit{time-dependent dielectric breakdown} (TDDB) \cite{sanyo},  and the \textit{negative-bias temperature instability} (NBTI).
All these permanent effects lead to increased transistor switching delays which further may lead to timing errors. A timing error occurs when the delay of a combinatorial path is increased so much due to the degeneration of corresponding path transistors that the clock period is exceeded. According to \cite{Stott:2010} and \cite{Stott:2010FPL}, the dominant degeneration effect in today's CMOS technologies is NBTI. Therefore, we focused on the NTBI effect in our work \cite{DBLP:conf/fpl/AngermeierZGT11} and \cite{DBLP:conf/fpt/AngermeierZGT11}. The NBTI effect depends mainly from the local temperature and gate voltage. The local temperature inside the FPGA depends on the local power consumption, the thermal resistance of the FPGA (e.g., package material, heat sink), and  the environment temperature.

The goal of our approach in \cite{DBLP:conf/fpl/AngermeierZGT11} was to determine the degradation of a reconfigurable device at run time. For this purpose, the additional delay of the individual reconfigurable resource due to aging should be estimated as accurate as possible under these circumstances. 
Due to the fact that temperature is the main accelerator for degeneration effects, the temperature profile of each active module is utilized to predict the degradation of each part of the reconfigurable area. Furthermore, this work proposed a stress-aware placing algorithm that predicts the degeneration based on an NBTI fault model, and aims at equally distributing modules to minimize the degradation.

\begin{figure}[t]
\centering
\includegraphics[width=340pt]{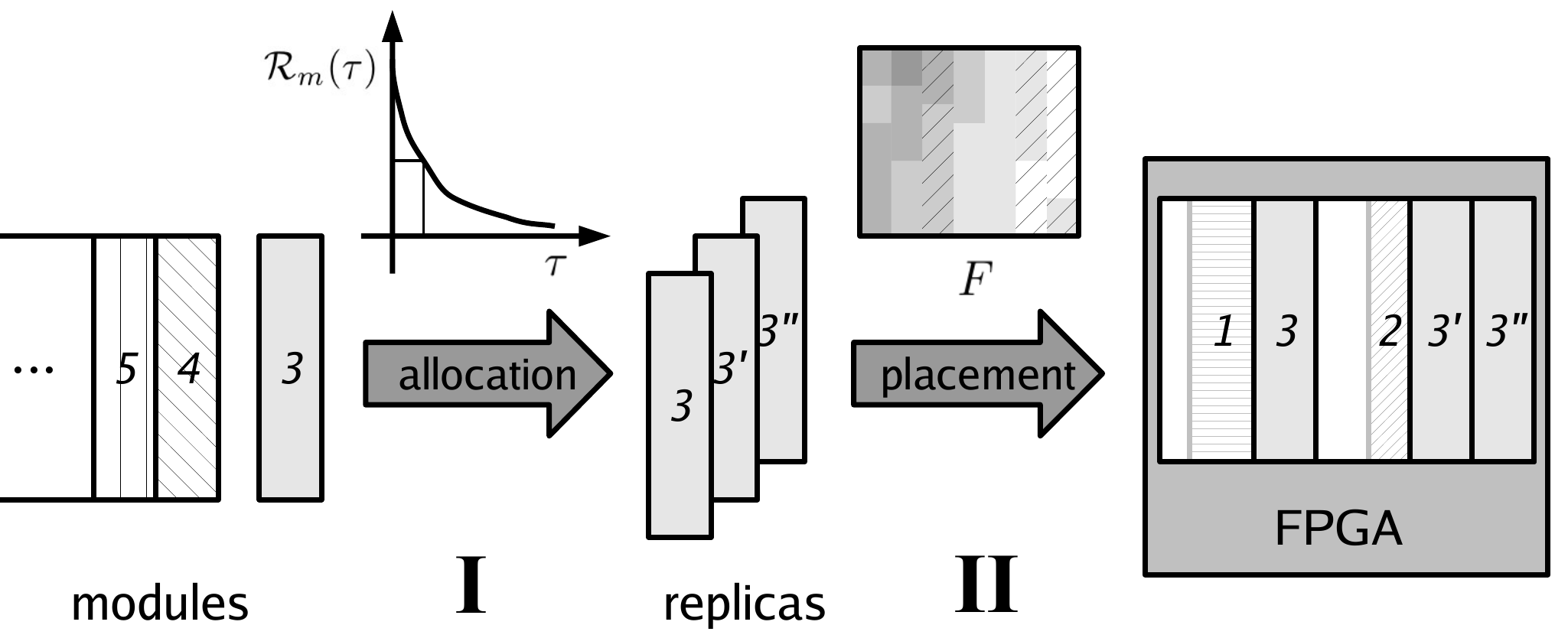}
\caption{A schematic overview of the proposed approach:
(I) Depending on the run time and number of sensitive configuration bits of a module, the number of replicas required to fulfill the respective safety level is determined.
(II) The replicas are placed according to a degeneration map $F$ of the reconfigurable area on the FPGA to minimize the maximum stress on the FPGA.   Taken from \cite{DBLP:conf/fpt/AngermeierZGT11}.
}
\label{fig:reoverview}
\end{figure}

The extension in \cite{DBLP:conf/fpt/AngermeierZGT11} introduced a two-step online approach for dynamically reconfigurable platforms that is capable of meeting safety requirements of modules by \emph{dynamically allocating replicas}. However, the allocation of replicas may sacrifice  the lifetime of the FPGA itself since it accelerates aging by increasing stress. The presented extended stress-aware placing algorithm then equally distributes active modules to minimize the degradation effects while respecting placement constraints that arise from the need for majority voting between the different replicas of a module (see Figure \ref{fig:reoverview}).

\subsection{Results}


A case study demonstrated that the proposed online placing algorithm can reduce the maximal degradation values by up to about $25 \%$, thus extending the overall lifetime of the reconfigurable devices by distributing the overall stress while, at the same time, hardening the system against radiation effects by instantiating replicas according to the required safety levels.

\subsection{Key Paper} 

I briefly describe the related key paper for this topic, which is
part of this cumulative habilitation treatise. Reprint of this paper is available in
Appendix~\ref{sec:mainpapers}.


\renewcommand{\arraystretch}{3.6}
\begin{tabular}{l l} 
 \pbox{4cm}{\textbf{FPT'11}} & \pbox{11cm}{\small Josef Angermeier, Daniel Ziener, Michael Gla{\ss},  and J{\"u}rgen Teich.  \emph{Runtime Stress-aware Replica Placement on Reconfigurable Devices under Safety Constraints} \cite{DBLP:conf/fpt/AngermeierZGT11}} \\
 &  \pbox{11cm}{ \vspace{5mm} This paper describes the modeling of aging effects for FPGA-based systems and the combination with an adaptive approach for replica placement against single event effects for space missions. The approach first allocates replicas of modules to cope with soft-errors and fulfill the required safety-level of each module.
Afterwards, it places the modules onto the FPGA in such a way that stress is minimized and, hence, degeneration of the device is reduced.
The number of required replica is determined at run time by applying a lifetime analysis of the reliability of a module depending on its sensitive configuration bits and the expected run time of the module.  \\
%
%
Furthermore, the degradation due to aging effects of each part of the reconfigurable area is estimated at run time by analyzing resource use and temperature distribution. Based on this information, the presented algorithm places active modules such that the degradation, i.\,e. increase in gate delay, is equally distributed on the whole reconfigurable device. \\
My personal contribution to this work was beside developing the basic concept, the calculation of the reliability values, the (co-) su\-per\-vision of the PhD student,  as well as writing around 40\% of the publication. } \\

 \end{tabular}
 
 \subsection{Team}

Publications \cite{DBLP:conf/fpl/AngermeierZGT11} and \cite{DBLP:conf/fpt/AngermeierZGT11} was conducted with Josef Angermeier, a former PhD student in my group.

\chapter{Improving Security by using Dynamic Hardware Reconfiguration} \label{sec:security}

In this chapter, the ideas to improve the security of implementations of cryptographic algorithms on reconfigurable systems by using dynamic partial reconfiguration are presented. These ideas lead to a successful project proposal. The project \emph{Security by Reconfiguration (SecRec)} funded by the Federal Ministry of Education and Research (BMBF) has started in January 2017. Project partners are  the \emph{DFKI} in Bremen, the \emph{Robert Bosch GmbH}, \emph{FZI Forschungszentrum Informatik} as well as the \emph{Mixed Mode GmbH}.  My subproject with two new doctoral positions for three years has the title \emph{Security by Reconfiguration -- Measures against Reverse Engineering and Fault Injection Attacks}. 

\section{Security by Reconfiguration (SecRec)}

Physical attacks such as side-channel analysis (a), fault injection attacks (b) or classical reverse-engineering (c) pose a massive threat to any cryptographic implementation. It is essential for any cryptographic hardware implementation that might be exposed to such physical attacks to include efficient countermeasures. 

FPGAs represent an efficient platform for cryptographic hardware implementations. However, in order to be hardened against any physical attacks, each security-critical circuit implemented on an FPGA must be protected against the above mentioned threats. In SecRec, three renowned research institutes and universities, one medium-size company with excellent credentials and one worldwide leading technology corporation are working together in order to jointly develop innovative approaches for hardware security implementations. In details, the SecRec project aims to develop techniques that are able to utilize the dynamic reconfiguration capabilities of FPGAs for effective protection mechanisms against the above mentioned class of attacks. The overall project will solve the fundamental problem of cryptographic hardware implementations, namely that these implementations exist only as static circuits and can therefore easily be analyzed. 

\subsection{Goals}

The primary goal of this project is to investigate to what extent the dynamic reconfiguration can be used as countermeasure against side-channel analysis, fault injection attacks, or classical reverse-engineering. In particular, this will address the problem that cryptographic hardware implementations are generally built by static circuits which even are shared by many instances and devices. Hence, an attacker can easily analyze one device to identify the internals and to acquire the knowledge to finally successfully mount an attack against the complete cryptographic system. In particularly, the subproject \emph{Security by Reconfiguration -- Measures against Reverse Engineering and Fault Injection Attacks} focuses on reconfiguration techniques in order to prevent fault inject attacks and reverse engineering.

\subsection{Problem Statement}

FPGAs are very suitable to host implementations of cryptographic algorithms, like AES \cite{aes} or RSA \cite{rivest1978method}. One reason is the efficient implementation of such streamed-based algorithms on FPGAs (see Chapter \ref{sec:efficiency}).  On the other hand, the possibility to reconfigure and update the implementation in the application field  after the  detection of security flaws is very valuable. However, the deployment of security relevant applications is not always in a trusted and controlled environment. It is rather so that the FPGA-based implementation is exposed to an environment which is under fully control of an attacker, who can launch physical attacks like manipulating clock, voltage, using radiation and laser beams and so on. These attacks are very powerful and almost all implementations suffer from these kinds of attacks. Therefore, measurements to protect the implementation for physical attacks are very important and should be carried out at run time by the FPGA platform itself. 

\subsubsection{Side-Channel Analysis}

Information of cryptographic operations in embedded systems can be gathered by side-channel analysis (SCA). Usually, the goal is to get the secret key. 
The different side-channel analysis are \textit{timing analysis} \cite{ko96}, \textit{power analysis} \cite{koc99}, and \textit{electromagnetic analysis} \cite{Agrawal2002} (see also \cite{disc, klm04, rr04}).   

Timing analysis attacks are based on the correlation of output data timing behavior and internal data values. Kocher  \cite{ko96} showed that it is possible to determine secret keys by analyzing small variations in the execution time of cryptographic computations. Different key bit values cause different execution time, which makes a read out and  reconstruction of the key possible. 

Power analysis attacks are based on the fact that different instructions cause variations in the activities on the signal lines, which result in different power consumption. The power consumption can be easily measured by observing the current or the voltage on a shunt resistor. With \textit{simple power analysis} (SPA) \cite{koc99}, the measured power consumption is directly interpreted to the different operations in a cryptographic algorithm. 
\textit{Differential power analysis} (DPA) \cite{koc99} is an enhanced method which uses statistical analysis, error correction and correlation techniques to extract exact information about the secret key. 
Similar to power analysis techniques, information about the key, data, or the cryptographic algorithm or implementation can be extracted by electromagnetic radiation analysis \cite{Agrawal2002}.  

Countermeasures to side-channel analysis can be categorized into algorithmic measures, like introduction of additional delays, shuffling of instructions, or masking of intermediate data with random values, and layout measures, like special logic gate styles  \cite{Tiri2002,Mace2004}. The latter approaches are clearly not applicable for reconfigurable devices. However, special countermeasures for reconfigurable systems using FPGAs resources  also exist \cite{IC11:Generic_FPGA-SCA}.

\subsubsection{Fault Injection Analysis}

During Fault Injection Analysis (FIA), the system is exposed to harsh environmental  conditions, like heat, vibrations, pressure, or radiation, to enforce faults which result in an erroneous output. Comparing this output to the correct one, the analyst gains insight into the implementation of the cryptographic algorithms as well as the secret key. With different fault analysis (DFA) attacks, it was possible to extract the key from a DES implementation \cite{bi97} as well as public key algorithm implementations \cite{bao98,bi97}. The last one shows that a single bit fault can cause fatal leakage of information which can be used to extract the key. Pellegrini and others show that using fault analysis, where the supply voltage of an processor is lowered, it is possible to reconstruct several bits from a secret key of the RSA cryptographic algorithm \cite{pba10}.  For this attack, they used the RSA implementation of the common OpenSSL cryptographic library and a SPARC Leon3 core, implemented on a Xilinx Virtex-II Pro FPGA.  The supply voltage of the FPGA is lowered till sporadic bit errors occur on the calculation of the signature using the FPGA's hardcore multiplier. Furthermore, the device can be exposed to a tightly focused laser beam \cite{canivet2011glitch}. Here, bit flips in registers and memory cells can be generated.    
Glitch attacks also belong to the class of FIA attacks. Here, additional glitches are inserted into the clock signal to prevent the registering of signal values on the critical path \cite{kom99}.  

Fault injection analysis for reconfigurable devices can be categorized into general fault injection attacks which are almost the same as in ASIC devices, and fault injection attacks on the configuration memory. Here, the configuration of different FPGA resources can be manipulated by altering the corresponding configuration bits. However, the corresponding configuration bits must be known beforehand (see reverse engineering attacks). Moreover, faults can also be injected by modifying  bits in an encrypted bitfile. In this case, the faults are not focused on a specific FPGA resource.  Countermeasures are usually redundancy and scrubbing  (see Chapter \ref{sec:relia}).  However, these measures are very costly in terms of resources and energy. 

\subsubsection{Reverse Engineering}

Reverse engineering is the analysis of the hardware design in order to detect and extract security elements embedded in hardware, e.g., static keys or security details of cryptographic implementations. A potential attacker might extract details from hardware circuits and convert it back to a netlist in order to analyze it and bringing the extracted component into semantical context.   The goal is to identify the functionality of cryptographic methods which is unknown for the attacker, or  to extract static integrated key fragments. This costly process which can only partly be done automatically, is, however, interesting for funded organizations like secret services. 

The general analysis of bit positions in FPGA configurations and the transfer to a simple netlist is shown in \cite{note2008bitstream}. Techniques to identify simple modules, like adders or multiplexers, from gate netlists are described in \cite{white2000candidate,doom1998identifying}. Even if these techniques do not describe a holistic reverse engineering approach, their potential to compromise security-related implementation of cryptographic components on such FPGA platforms is obviously a risk that should not be underestimated.

To protect FPGA-based implementations from these kinds of attacks, FPGA vendors introduced the encryption of the FPGA \textit{configuration} or \textit{bitfile} with symmetrical cryptographic approaches, like AES.  The bitfile is stored in a \textit{non-volatile memory}, e.g., a PROM, and transferred to the FPGA encrypted. Inside the FPGA during the configuration, the bitfile is decrypted.  However, current implementations of bitstream decryption have weaknesses which can be exploited by side-channel analysis \cite{DBLP:conf/fpga/MoradiOPS13}. Another approach which can also be used additionally to bitstream encryption is hardware obfuscation which can be used on RTL level \cite{brzozowski2007obfuscation,chakraborty2010rtl} as well on logic level \cite{mcdonald2009protecting}.

\subsection{Approach}

Analyzing the basic problem of physical attacks in details, it becomes clear that the basic principle of cryptography (\emph{Confidentiality}) cannot be transferred to the underlying circuit. An attacker with physical access to the device has no technical and time limits to identify and characterize the security functions and used cryptographic circuits. This is favored by the fact that cryptographic hardware implementations are present as \emph{static circuits}, which are usually identical over many device instances. Thus the attacker can carry out structural investigations completely destructively and temporally decoupled by means of several (similar) devices or combined attack techniques.

Conversely, all physical attacks on cryptographic hardware implementations are possible because their circuit structures are static and can, thereby, be systematically measured and manipulated. If, however, it is possible to continually change such a hardware configuration by means of reconfigurable technology in an invariable and dynamic manner at run time, the static attack principles of the SCA, FIA and the RE would certainly render a physical characterization by an attacker difficult or even impossible. In such a restructuring of the circuit, the time points and the location of the processing of sensitive information could be shifted during each cryptographic operation in such a way that an attacker is prevented from creating a suitable attack model and characterizing internal processes. A first approach in this direction  to replace parts of a cryptographic implementation at run time by means of partial reconfiguration is proposed in \cite{IC08_Mentens}. However, it is neither systematically investigated which parts are particularly suitable for this purpose, nor did the proposed initial concept provide sufficient dynamics and complexity to provide a significant protection.

The central goal is thus to investigate how cryptographic implementations can continuously evolve in hardware by (1) implementing local modifications of lookup tables or memory blocks or (2) by use of the partial dynamic reconfiguration during run time. Local reconfiguration replaces configuration of special FPGA elements, whereas partial dynamic reconfiguration enables a replacement of complete sub circuits at run time. 
Although the process of dynamic reconfiguration of hardware circuits certainly requires a significant amount of resources compared to its purely static implementation, the approach still appears extremely promising since it combines countermeasures against all of the above mentioned attacks at the same time. In particular, a large number of individual countermeasures have been combined until now to thwart all types of physical attacks. This approach obviously also introduces significant costs and increased run time typically resulting in an overhead of a factor of 2-3 orders of magnitude. 


 
\paragraph{Protection by Local Reconfiguration}
 
Since current FPGAs can efficiently modify the configuration of their lookup tables and block RAMs at run time, the protection through local reconfiguration examines how basic cryptographic functions of symmetric and asymmetric cryptographic algorithms can be modified at run time. Since the routing of reconfigurable circuits can only be changed to a very limited extent at run time, time-division multiplexing methods or \emph{shuffling}~\cite{Mangard2007} by shift register LUTs (SRL) will be evaluated.

\paragraph{Protection by Partial Reconfiguration}

The partial reconfiguration can be used to configure the FPGA at run time with various implementations of an identical function. These implementations could also be generated during run time. At random time points of a cryptographic operation, individual parts of the implementation are completely exchanged and reloaded. Compared to the local reconfiguration, the partial reconfiguration is a much more complex process. However, it offers the possibility to rearrange all parts (i.e., routing and logic elements) in a completely new way. As a drawback,  an increased reconfiguration time as well as a temporary interruption of switching parts must  be taken into account.
 
\subsection{Supervised Theses}

In this research area, I (co-)supervised the following theses:

\begin{itemize}

\item Christian Hilgers, Masterarbeit, \emph{Sicheres Laden und Authentifizieren von partiellen Modulen f{\"u}r 						dynamisch rekonfigurierbare SoCs} (engl. Secure Loading and Authentication of Partial Modules for Dynamically Reconfigurable SoCs), Hardware/Software Co-Design, FAU Er\-lang\-en-N{\"u}rnberg, September 2014

\item Michael Weisser, Masterarbeit, \emph{Seitenkanalangriffe auf FPGA-basierten kryptographischen 						Algorithmen}  (engl. Side Channel Attacks on FPGA-based Cryptographic Algorithms), Hardware/Software Co-Design, FAU Er\-lang\-en-N{\"u}rnberg, March 2015

\item Maike Monika Meier, Projektarbeit, \emph{Anaylsis of Fault Injection Attacks on FPGA-based Cryptographic Implementations}, Institute of Embedded Systems, TU Hamburg, February 2017 

\end{itemize}

\chapter{Conclusions \& Future Work} \label{sec:con}

This treatise summarizes my research on improving efficiency, reliability, and security of reconfigurable systems. The improvements in these important properties pave the way for extending the application field for embedded systems in general and also for hardware accelerators in big data applications. Our approach to use area and energy-efficient hardware designs and turning them from pure static implementations to dynamic adaptive systems has been successfully demonstrated.  The possibility to adapt the structure of a hardware system during run time was shown by using partial reconfiguration of FPGAs. This feature of current FPGAs allows to build very flexible and efficient systems. By combining the efficiency of hardware-based systems with the flexibility of software-based systems, these adaptive systems  open up the way to new application fields for reconfigurable systems. The purchase of the major FPGA vendor Altera was the biggest acquisition of  Intel. This shows that the combination of reconfigurable systems with high-end processors is seen as a big chance to increase performance and efficiency not only for small embedded devices, but also for data centers with big data applications. Energy-aware processing is utmost important for future data center applications.  The energy consumption of all IT facilities together approached 10\% of the worldwide energy generation \cite{coal} in 2013, and it will increase further. For example, a single Google search inquiry requires around 0.3 Wh which is the equivalent of switching on a non-environment-friendly  60 W light bulb for 18 seconds \cite{googleenergy}. This development forces us to investigate more in efficient data processing approaches such as the approaches shown here in the treatise. 

Furthermore, I showed in this treatise that the reliability against radiation-induced faults and aging effects could be significantly increased by using such adaptive systems. Moreover, I presented ideas to increase also the protection against physical attacks on embedded systems by using adaptive circuits. This will be one of my important areas of my future research. 

\section{Future Work}

The combined energy-efficient hardware acceleration of different data-intensive  tasks, like de- and encryption of stored data or data to transmit, query processing and filtering as well as data analytics, e.g., with hardware machine learning techniques, in a general adaptive data acceleration platform for data-intensive systems would be an interesting future research topic. Beside the envisaged accelerators, also the management of these accelerators is important. Research questions in this area that I would like to address  are: How can the available accelerators be shared for many user requests? How can the response time for requests be lowered without introducing additional overheads? What are the best storage architectures to avoid costly data transmissions? How can the data integrity and security of  different tasks on the same accelerator platform be ensured? To circumvent the I/O bottleneck for processor attached accelerators,  \emph{solid state disks} (SSDs) could be directly attached to the FPGA. This strategy exploits synergism of low power, low footprint, and high storage capacity. The usage of high throughput compression and decompression modules inside the FPGA increases the bandwidth between the storage and the FPGA further. 

Furthermore, hardware acceleration for high performance computing as well as  image and media processing might also be valuable to investigate. These energy-efficient data processing techniques are essential for the envisaged ubiquitous computing platforms which especially need to deal with limited resources. Furthermore, research on approximate computing for FPGAs is also an important research topic which can be combined with the research on accelerators. 

Moreover, asynchronous systems are also very interesting in the security area for improving the robustness of cryptographic implementations against side-channel attacks.  The data-dependent timing might help to introduce jitters on data processing. On the other hand, it could also lead to fatal information leak which could  be exploited by attacks. However, by combining such asynchronous systems with dynamical reconfigurable systems, introduced in this treatise, the information leak could be masked and to the attack complexity could be enormously increased.


\clearpage
\appendix
\include{app_assessment}
\chapter{Bibliography}

\nocite{patep,patus}
\nocite{ZBB16,WRZT:2013,ZT09,ZT07,KHZ16,HKZ15,HKZ14,zie10,KZH16,zst10,EWBTZ16, PZ6, BEZWT:2016, BZMT15,BEZT15,GRBZFTH15,DBLP:conf/fpl/Becher14, RAW:2014, FCCM:2014, Schmidt:2014:ANF:2554688.2554730, SZT14,FCCM:2013,  ZBTZ12,GYRT2012, FCCM:2012, KZF12,DBLP:conf/fpt/AngermeierZGT11,DBLP:conf/codes/WildermannRZT11,DBLP:conf/fpl/AngermeierZGT11, DBLP:conf/fpl/WildermannTZ11, ZWOWT11, TZ11, ZT10,ZBT10b, ZBT10,
 MBZ10,SZT2008,ZT2008, AISEC2007, AIS2007, ZT2006, ZAT2006,AHWZ06, DBLP:conf/fpt/ZiermannSMZAT11,AIS2009,ZT05}

\section{General Bibliography}
\newrefcontext[sorting=nyt]
\printbibliography[notcategory=patents,notcategory=journals,notcategory=books,notcategory=bookchapters,notcategory=conferences,notcategory=miscs,heading=subbibliography,title={General Bibliography},heading=none]

%
%
%
%
%
%
%
 \cleardoublepage

\section{Personal Bibliography}\label{sec:mypup}

\let\etalchar\undefined 

Note that the publications included in this cumulative treatise in Appendix B are highlighted with blue color. 

\subsubsection{Awards:}

\begin{tabbing}
\hspace{2cm} \= \hspace{7cm} \= \kill
\>Best Paper Award: \> FPT 2006, ReConFig 2016 \\
\>Best Application Paper Award: \> AHS 2015 \\
\>Best Paper Candidate: \> CODES+ISSS 2011 \\ 
\end{tabbing}

\subsubsection{Patents:}
\newrefcontext[labelprefix=P]
\printbibliography[category=patents,resetnumbers,heading=none]
 
%

%
%
%
%
%
%
%
%
%
\subsubsection{Books:}

\newrefcontext[labelprefix=B]
\printbibliography[sorting=ydnt,category=books,resetnumbers,heading=none]

%
%
%
\subsubsection{Book Chapters:}

\newrefcontext[labelprefix=BC]
\printbibliography[sorting=ydnt,category=bookchapters,resetnumbers,heading=none]

%
\subsubsection{Journals:}

\newrefcontext[labelprefix=J]
\printbibliography[sorting=ydnt,category=journals,resetnumbers,heading=none]

%

%
\subsubsection{Conferences: (peer reviewed)}

\newrefcontext[labelprefix=C,sorting=ydnt]
\printbibliography[category=conferences,resetnumbers,heading=none]

%
%
%
%
 \subsubsection{Miscs:}
 
 \newrefcontext[labelprefix=M]
\printbibliography[sorting=ydnt,category=miscs,resetnumbers,heading=none]

%
%

\chapter{Paper Reprints}\label{sec:mainpapers}

This document is a cumulative habilitation treatise. I have selected out of my 45 peer-reviewed publications, listed in Appendix \ref{sec:mypup}, the following seven papers as the key contributions of my research. 

\vspace{0.5cm}

\noindent \textbf{Methods for Improving the Efficiency of Reconfigurable Systems:}

\vspace{-2mm}

\renewcommand{\arraystretch}{3.6}
\begin{tabular}{l l l} 
\textbf{FCCM'12} & \pbox{10cm}{\small Christopher Dennl, Daniel Ziener, and J{\"u}rgen Teich.  \emph{On-the-fly Composition of FPGA-Based SQL Query Accelerators Using A Partially Reconfigurable Module Library}} & \cite{FCCM:2012} \\
\textbf{FPL'14} & \pbox{10cm}{\small Andreas Becher, Florian Bauer, Daniel Ziener, and J{\"u}rgen Teich.  \emph{Energy-Aware SQL Query Acceleration through FPGA-Based Dynamic Partial Reconfiguration.}} & \cite{DBLP:conf/fpl/Becher14} \\
\textbf{FPT'15} & \pbox{10cm}{\small Andreas Becher, Daniel Ziener, Klaus Meyer-Wegener, and J{\"u}rgen Teich.   \emph{A Co-Design Approach for Accelerated SQL Query Processing via FPGA-based Data Filtering}} & \cite{BZMT15} \\
\textbf{TRETS'16} & \pbox{10cm}{\small Daniel Ziener, Florian Bauer, Andreas Becher, Christopher Dennl, Klaus Meyer-Wegener, Ute Sch{\"u}rfeld, J{\"u}rgen Teich, J{\"o}rg-Stephan Vogt, and Helmut Weber.  \emph{FPGA-Based Dynamically Reconfigurable SQL Query Processing}} & \cite{ZBB16} \\
\end{tabular}

\vspace{0.5cm}

\noindent \textbf{Methods for Improving the Reliability of Reconfigurable Systems:}
\vspace{-2mm}

\begin{tabular}{l l l} 

\textbf{FPT'11} & \pbox{10cm}{\small Josef Angermeier, Daniel Ziener, Michael Gla{\ss},  and J{\"u}rgen Teich.  \emph{Runtime Stress-aware Replica Placement on Reconfigurable Devices under Safety Constraints}} & \cite{DBLP:conf/fpt/AngermeierZGT11} \\
\textbf{FCCM'14} & \pbox{10cm}{\small Robert Glein, Bernhard Schmidt, Florian Rittner, J{\"u}rgen Teich, and Daniel Ziener.  \emph{A Self-Adaptive SEU Mitigation System for FPGAs with an Internal Block RAM Radiation Particle Sensor}} & \cite{FCCM:2014} \\
\textbf{AHS'15} & \pbox{10cm}{\small Robert Glein, Florian Rittner, Andreas Becher, Daniel Ziener, J{\"u}rgen Frickel, J{\"u}rgen Teich, and Albert Heuberger. \emph{Reliability of Space-Grade vs. COTS SRAM-Based FPGA in N-Modular Redundancy}} & \cite{GRBZFTH15} \\

\end{tabular}

\section{On-the-fly Composition of FPGA-Based SQL Query Accelerators Using A Partially Reconfigurable Module Library}

\cite{FCCM:2012} Proceedings of the IEEE International Field-Programmable Custom Computing Machines Symposium (FCCM), 2012


\vspace{1cm}

\emph{Note: Due to copyright reasons, only the URL to the original paper is included in this online version of this treatise.}

\vspace{1cm}

\url{http://ieeexplore.ieee.org/abstract/document/6239790/}

\newpage

\section[Energy-Aware SQL Query Acceleration through FPGA-Based Dynamic Partial Reconfiguration]{Energy-Aware SQL Query Acceleration through FPGA-Based Dynamic Partial Reconfiguration%
\sectionmark{Energy-Aware SQL Query Acceleration through FPGA-Based DPR}}
\sectionmark{Energy-Aware SQL Query Acceleration through FPGA-Based DPR}

\cite{DBLP:conf/fpl/Becher14} Proceedings of the International Conference on Field Programmable Logic and Applications (FPL), 2014


\vspace{1cm}

\emph{Note: Due to copyright reasons, only the URL to the original paper is included in this online version of this treatise.}

\vspace{1cm}

\url{http://ieeexplore.ieee.org/abstract/document/6927502/}

\newpage

\section{A Co-Design Approach for Accelerated SQL Query Processing via FPGA-based Data Filtering}

\cite{BZMT15} Proceedings of the IEEE International Field-Programmable Custom Computing Machines Symposium (FCCM), 2015


\vspace{1cm}

\emph{Note: Due to copyright reasons, only the URL to the original paper is included in this online version of this treatise.}

\vspace{1cm}

\url{http://ieeexplore.ieee.org/abstract/document/7393148/}

\newpage

\section{FPGA-Based Dynamically Reconfigurable SQL Query Processing}

\cite{ZBB16} ACM Transactions on Reconfigurable Technology and Systems (TRETS), 2016 


\vspace{1cm}

\emph{Note: Due to copyright reasons, only the URL to the original paper is included in this online version of this treatise.}

\vspace{1cm}

\url{https://dl.acm.org/citation.cfm?id=2845087}

\newpage

\section{Runtime Stress-aware Replica Placement on Reconfigurable Devices under Safety Constraints}

\cite{DBLP:conf/fpt/AngermeierZGT11} Proceedings of the International Conference on Field-Programmable Technology (FPT), 2011


\vspace{1cm}

\emph{Note: Due to copyright reasons, only the URL to the original paper is included in this online version of this treatise.}

\vspace{1cm}

\url{http://ieeexplore.ieee.org/abstract/document/6133247/}

\newpage

\section[A Self-Adaptive SEU Mitigation System for FPGAs with an Internal Block RAM Radiation Particle Sensor]{A Self-Adaptive SEU Mitigation System for FPGAs with an Internal Block RAM Radiation Particle Sensor%
\sectionmark{A Self-Adaptive SEU Mitigation System for FPGAs}}

\cite{FCCM:2014} Proceedings of the IEEE International Field-Programmable Custom Computing Machines Symposium (FCCM), 2014


\vspace{1cm}

\emph{Note: Due to copyright reasons, only the URL to the original paper is included in this online version of this treatise.}

\vspace{1cm}

\url{http://ieeexplore.ieee.org/abstract/document/6861641/}

\newpage

\section{Reliability of Space-Grade vs. COTS SRAM-Based FPGA in N-Modular Redundancy}

\cite{GRBZFTH15} Proceedings of 2015 NASA/ESA Conference on Adaptive Hardware and Systems (AHS), 2015


\vspace{1cm}

\emph{Note: Due to copyright reasons, only the URL to the original paper is included in this online version of this treatise.}

\vspace{1cm}

\url{http://ieeexplore.ieee.org/abstract/document/7231159/}

\newpage

\backmatter
\selectlanguage{english}

\cleardoublepage
\printindex
\cleardoublepage
\end{document}